\newcount\draft\draft=0 
\newcount\cameraready\cameraready=0

\documentclass[acmsmall,screen]{acmart}
\renewcommand\footnotetextcopyrightpermission[1]{} 
\usepackage{xxx}
\usepackage{color}
\usepackage{multirow}

\DeclareGraphicsExtensions{.pdf,.jpg,.png}
\graphicspath{{./figs/} {./plots/}}
\usepackage{enumitem}
\usepackage{wrapfig}
\usepackage{algorithm}
\usepackage[noend]{algpseudocode}
\usepackage{subcaption}
\usepackage{xspace}
\usepackage{makecell}

\newcommand{\sys}{Alpaca\xspace}
\newcommand{\dino}{DINO\xspace}
\newcommand{\chain}{Chain\xspace}
\newcommand{\ratchet}{Ratchet\xspace}

\newcommand{\trans}{{\tt transition\_to}\xspace}
\newcommand{\dirtylist}{\texttt{commit\_list}\xspace}
\newcommand{\war}{\textit{W-A-R}\xspace}
\newcommand{\precommit}{\textit{pre-commit}\xspace}

\newcommand{\precommitfunc}{\texttt{pre\_commit}\xspace}
\newcommand{\commit}{\textit{commit}\xspace}
\newcommand{\commitfunc}{\texttt{commit}\xspace}

\newcommand{\numBoots}{{\tt cur\_version}\xspace}
\newcommand{\task}{{\tt task}\xspace}
\newcommand{\commitready}{{\tt commit\_ready}\xspace}
\newcommand{\commitendindex}{{\it end-index}\xspace}
\newcommand{\taskshared}{{\tt TS}\xspace}
\newcommand{\entrytask}{{\tt entry}\xspace}
\newcommand{\initfunc}{{\tt init}\xspace}

\begin{document}


\title{Alpaca: Intermittent Execution without Checkpoints}


\author{Kiwan Maeng}
\affiliation{
  \department{Electrical and Computer Engineering Department}              
  \institution{Carnegie Mellon University}            
  \country{USA}
}
\email{kmaeng@andrew.cmu.edu}          

\author{Alexei Colin}
\affiliation{
  \department{Electrical and Computer Engineering Department}              
  \institution{Carnegie Mellon University}            
  \country{USA}
}
\email{acolin@andrew.cmu.edu}          

\author{Brandon Lucia}
\affiliation{
  \department{Electrical and Computer Engineering Department}              
  \institution{Carnegie Mellon University}            
  \country{USA}
}
\email{blucia@andrew.cmu.edu}          




\begin{abstract}
The emergence of energy harvesting devices creates the potential for
batteryless sensing and computing devices. Such devices operate only
intermittently, as energy is available, presenting a number of challenges
for software developers.  Programmers face a complex design space
requiring reasoning about energy, memory consistency, and forward progress.
This paper introduces \sys, a low-overhead programming model for intermittent
computing on energy-harvesting devices.  \sys programs are composed of a
sequence of user-defined tasks. The \sys runtime preserves execution progress
at the granularity of a task.  The key insight in \sys is the {\em
privatization} of data shared between tasks.
Updates of shared values in a task are privatized and
only committed to main memory on successful execution of the task,
ensuring that data remain consistent
despite power failures.  \sys provides a familiar programming interface and a
highly efficient runtime model.
We also present an alternate version of \sys, Alpaca-undo, that uses
undo-logging and rollback instead of privatization and commit.
We implemented a prototype of both versions of \sys as an extension to C with an LLVM compiler
pass. We evaluated \sys, and directly compared to three systems from prior work.
\sys consistently improves performance compared to the previous systems, by up to 23.8x, while
also improving memory footprint in many cases, by up to 17.6x.  

\end{abstract}




\maketitle

\section{Introduction}
\label{sec:intro}

The emergence of extremely energy-efficient processor architectures creates the
potential for computing and sensing systems that operate entirely using energy
extracted from their environment.  Such {\em energy-harvesting} systems can use
energy from radio waves~\cite{wisp,moo}, solar energy~\cite{kicksat,m3}, and
other environmental sources.  An energy-harvesting system operates only {\em
intermittently} when energy is available in the environment and experiences a
power failure otherwise.  To operate, a device slowly buffers energy into a
storage element (e.g., a capacitor).  Once sufficient energy accumulates, the
device begins operating and quickly consumes the stored energy.  Energy
depletes more quickly during operation (e.g. milliseconds) than it accumulates
during charging (e.g., seconds).  When energy is depleted and the device powers
off, volatile state, e.g. registers and stack memory, is lost, while
non-volatile state, e.g., ferroelectric memory (FRAM), persists.
The charge/discharge cycle of an energy-harvesting device forces software to
execute according to the {\em intermittent execution model}~\cite{dino, chain, ratchet}.
An intermittent execution includes periods of activity perforated
by power failures.  The key distinction between intermittent execution and
continuously-powered execution is that in the intermittent model a computation
may execute only partially before power fails and must be resumed after the
power is restored.
Correct and efficient intermittent execution requires a system to meet a 
set of correctness requirements (C1-3) and performance goals (G1-3).

\begin{enumerate}[label={\bf C\arabic*:}]
\item{A program must preserve progress despite losing volatile state on power failures.} 
\item{A program must have a consistent view of its state across volatile and non-volatile memory.}
\item{A program must respect atomicity constraints (e.g., sampling related sensors together).}
\end{enumerate}

\begin{enumerate}[label={\bf G\arabic*:}]
\item{Applications should place as few restrictions on the hardware as possible.}
\item{Applications should be tunable at design time to use the energy storage capacity efficiently.}
\item{Applications should minimize runtime overhead and memory footprint.}
\end{enumerate} 
 
Recent work made progress toward several of these goals, but 
necessarily compromised on others.
This paper develops \sys~\footnote{Alpaca: Adaptive Lightweight Programming
Abstraction for Consistency and Atomicity}, a programming and execution model
that allows software to execute intermittently.  Like state-of-the-art
systems, \sys preserves progress despite power failures (C1) and ensures memory
consistency (C2).  
\sys uses a {\em static task model} that can adhere to programmer-provided atomicity
constraints and energy availability (G2, C3).
Memory updates made by an \sys task only commits atomically when the task completes.
By discarding memory updates on power failure, \sys can restart a task with negligible cost,
without checkpointing the volatile state as in prior work~\cite{dino, ratchet, mementos} (G3).
Unlike prior work that requires the entire memory to be non-volatile~\cite{ratchet},
\sys can leverage both volatile and non-volatile memory (G1).
We present two different versions of Alpaca with different design choices.
\sys's design differences, relative to state-of-the-art systems, translate into
performance gains of 4-5.2x on average (up 23.8x in some cases). Also, \sys shows
smaller memory footprints compared to most of the previous systems.
%

Section~\ref{sec:background}  provides background on intermittent computing.
Sections~\ref{sec:alpacalang} and \ref{sec:impl} describe the \sys programming
model and its implementation. Section~\ref{sec:undo} presents an alternate design
choice of \sys. Section~\ref{sec:discussion} discusses key design
decisions.  Sections~\ref{sec:apps} and \ref{sec:eval} describe our benchmarks
and evaluation. We conclude with a discussion of related 
(Section~\ref{sec:related}) and future (Section~\ref{sec:conclusion}) work.

\section{Background and Motivation}
\label{sec:background}

Energy-harvesting systems operate intermittently, losing power frequently and
unexpectedly.  Intermittent operation compromises forward progress and leads to
inconsistent device and memory states, with unintuitive
consequences that demand abstraction by new programming models.

\subsection{Energy-Harvesting Devices and Intermittent Operation}

Energy-harvesting devices operate using energy extracted from their
environment, such as solar power~\cite{kicksat,m3}, radio waves
(RF)~\cite{wisp}, or mechanical interaction~\cite{papergenerators,mitbutton}.
As the processor on such a device executes software to interact with sensors
and actuators or communicate via radio, it manipulates both volatile and
non-volatile memory.  
An energy-harvesting device can operate only intermittently, when energy is
available.  Common energy-harvesting platforms~\cite{wisp} use a power system
that charges a capacitor slowly to a threshold voltage.  At the threshold, the
device begins operating, draining the capacitor's stored energy much more
quickly than it can recharge. The system eventually depletes the capacitor, and
the device turns off and waits to again recharge to its operating voltage.
These power cycles can occur frequently: RF-powered devices may reboot
hundreds of times per second~\cite{wisp}. 

\subsection{Device Model and Hardware Assumptions} 
Our work makes few
assumptions about device hardware. A device's memory system can include an
arbitrary mixture of volatile and non-volatile memory, unlike prior work that
requires all memory to be non-volatile~\cite{nvp,ratchet,quickrecall}. \sys
works on devices with non-volatile memories that support atomic read and write operations,
e.g. Ferroelectric RAM~\cite{catalog} and Flash. In commercially
available FRAM implementations that rely on destructive reads (i.e.,
rewrite-on-read), access atomicity is satisfied by means of an internal
capacitor that buffers sufficient energy to complete an in-progress access.  Our model allows arbitrary peripheral (I/O) devices as
detailed in Section~\ref{sec:discussion}.

\vspace{-5pt}
\subsection{Intermittent Execution and Memory Consistency}
\label{sec:background:consistency}

\begin{wrapfigure}{t}{0.46\textwidth}
	\centering
	\begin{subfigure}[t]{0.45\textwidth}
		\centering
		\includegraphics[width=\textwidth]{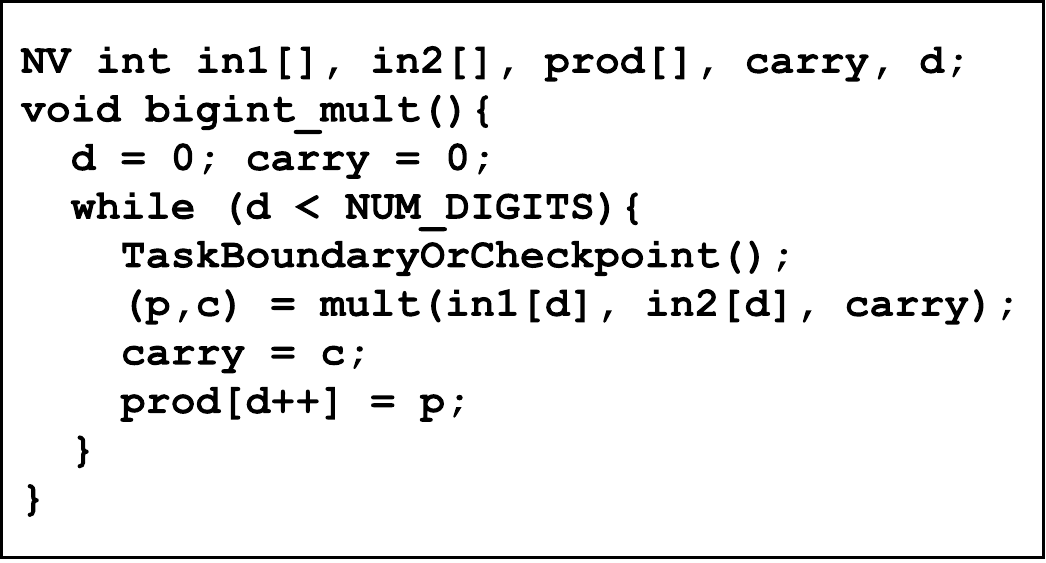}
		\caption{{Sample code from RSA.}}
		\label{fig:rsawarcode}
	\end{subfigure}
	\begin{subfigure}[t]{0.45\textwidth}
		\centering
		\includegraphics[width=\textwidth]{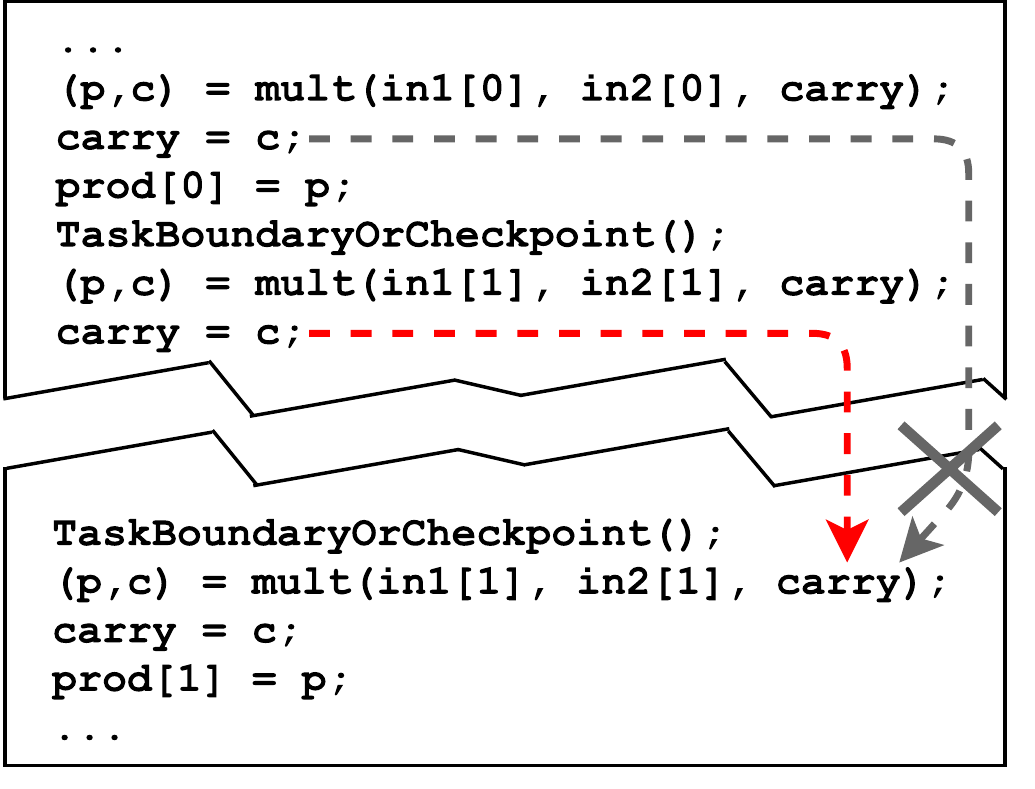}
		\caption{{Intermittent execution.}} 
		\label{fig:rsawarexec}
	\end{subfigure}
	\caption{{RSA code with intermittent execution.}}
	\label{fig:rsa}
\end{wrapfigure}

Software on an energy-harvesting device operates {\em intermittently}:  an
intermittent execution does not {\em end} when power fails; instead the
execution alternates between active periods and inactive periods.  On each
power failure, the register file and volatile memory (i.e., stack and globals) are erased.
Variables in non-volatile memory persist.
Prior work~\cite{mementos,hibernus,hibernusplusplus,idetic,quickrecall}
checkpoints volatile state periodically and restores a checkpoint after a power
failure.   Other prior work~\cite{dino, ratchet, chain} found that if an
application directly manipulates non-volatile memory, checkpointing only the volatile
state is not enough to guarantee consistency.  The problem exists because
some memory operations may repeat after restarting from a checkpoint.
Non-volatile state written before a power failure persists after a restart, and
if re-executing code reads the non-volatile state without first over-writing
it, the code may operate using inconsistent values. The resulting program behavior is
impossible if the device were powered continuously.  Precisely, a non-volatile value that may be
read and later written (i.e., a ``write-after-read'', or \war) between two
consecutive checkpoints can become inconsistent~\cite{ratchet,idempotent,dino}.

Figure~\ref{fig:rsa} illustrates how the combination of a \war
dependence and volatile-only checkpointing can leave data inconsistent.  The
code, excerpted from our implementation of RSA~\cite{rsa}, multiplies two
numbers {\tt in1} and {\tt in2} digit by digit, accounting for carries.
A task boundary
or a checkpoint is denoted uniformly by \texttt{TaskBoundaryOrCheckpoint()}.
The {\tt NV} prefix denotes non-volatile
data.  The code preserves per-digit progress using non-volatile variables {\tt
d}, {\tt carry}, and {\tt prod[]}, the output digit index, most recent carry
value, and output product. In the execution, {\tt carry} is updated, power
fails, and after restarting, {\tt mult()} uses the already-updated value of
{\tt carry}, producing the wrong result (Figure~\ref{fig:rsawarexec}).  The code first reads, then writes
{\tt carry} (a \war), putting it at risk of inconsistency.  While the
figure shows a problem with {\tt carry} only, {\tt d} is also read,
then written, presenting another potential consistency problem.

\subsection{Overhead of Existing Approaches}

Intermittent programming systems that handle volatile and non-volatile memory
consistency preserve progress across power failure by either taking
checkpoints~\cite{ratchet} or bounding tasks~\cite{dino,chain}.
Compiler-automated checkpointing approaches~\cite{ratchet, mementos}
are limited to their static analysis, often resulting
in much frequent checkpointing then necessary~\cite{ratchet}.
They also often copies the entire memory, wastefully copying even what has not been
updated~\cite{mementos}, or copies much more than necessary due to the limitation
of pointer aliasing~\cite{dino}.
Moreover, they cannot adhere to high-level atomicity constraints.
System asking the programmer to place boundaries~\cite{dino, chain}
does not incur frequent checkpointing overhead. However, to make
memory consistent across power failure, previous work relied on a
custom data structure with high space overhead (i.e., a ``channel'')~\cite{chain},
or a compiler analysis which is prone to high overhead due to conservatism
of the static analysis~\cite{dino}.

\section{\sys Programming Model}
\label{sec:alpacalang}


\sys is a programming interface that allows programmers to write software that
behaves correctly under an intermittent execution model.  \sys aims to overcome
the limitations of prior work described in Section~\ref{sec:background} and to
meet design requirements C1--C3 and design optimization goals G1--G3 from
Section~\ref{sec:intro}. 
The \sys programming model consists of two core concepts, \textit{tasks} and
\textit{privatization}. A task is a programming abstraction that is useful for
preserving progress, implementing atomicity constraints, and controlling an
application's energy requirements. Privatization is a language feature that
guarantees that any volatile or non-volatile memory accessed
by a task remains consistent, regardless of power conditions.

%
%


\subsection{Task-Based Programming}
\label{sec:task-based}

A task in \sys is a user-defined region of code that executes on a consistent
snapshot of memory and produces a consistent set of outputs. An \sys task that
eventually has sufficient energy to execute to completion is guaranteed to have
behavior (i.e., control-flow and memory reads and writes) that is equivalent to
some continuously-powered execution regardless of arbitrarily-timed power
failures.  As Section~\ref{sec:impl} describes, if
power fails during a task's execution, \sys effectively discards intermediate results and
execution starts again from the beginning of the task.  Consequently, a programmer can reason as though tasks are atomic,
like transactions in a TM system.  Computations that consume more energy than
the hardware can provide between two consecutive power failures must be split
into multiple tasks.

\begin{figure}[t]
	\centering
		\includegraphics[width=0.95\textwidth]{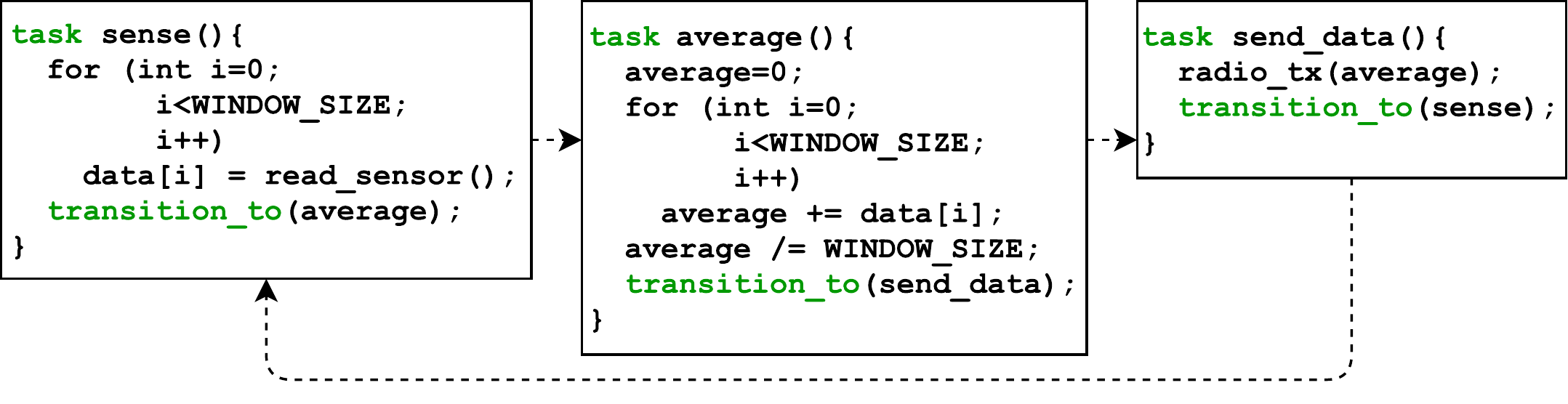}
    \caption{{An application written in \sys.} The program samples
    a sensor, calculates an average, and transmits via radio.} 
	\label{fig:example}
\end{figure}

To program in \sys, the programmer decomposes application code into tasks,
each marked with the \task keyword. 
Each task explicitly transfers control to another task (or to itself).
A program's control flow is defined by the execution of tasks in the sequence
specified by the transfer statements.
To transfer control from a task to one of its successors, the
programmer uses the \trans keyword, which takes the name of a task as its
argument and immediately jumps to the beginning of that task.  
\trans statements are valid along any control-flow path within a task, and all
paths through a task must end in a \trans statement or program termination.
The programmer specifies which task should run on when the system powers on
for the first time using {\tt entry} keyword. 
%
Figure~\ref{fig:example} shows a sensing application written using 
\sys.
\sys tasks are syntactically similar to Chain tasks~\cite{chain}, but the 
memory model for task interactions differs completely.

\sys guarantees to the programmer that a task executes {\em atomically} even
if power fails during its execution.  
When the task completes and the next task begins, changes to memory made by the
completed task are guaranteed to be visible and control never flows backward to
the completed task again, unless an explicit \trans statement executes.
Conversely, if a task does not complete due to a power failure, control does
not advance to any other task, which prevents the partially updated state from
becoming visible.
\sys allows only a single task sequence and does not support parallel
task execution.
This design choice is reasonable because parallel hardware is extremely rare
in intermittent devices due to its relatively high power consumption.  \sys does not
support concurrent (i.e., interleaved as threads) task sequencing. Concurrency
is limited to I/O routines only, which are addressed in Section~\ref{sec:io}.

Task atomicity guarantees correctness by ensuring that if any of a task
execution's effects become visible, then all of them are visible, and by ensuring that a
completed task's execution takes effect only once.  Moreover, task-based
execution {\em preserves progress}, assuming that eventually the system buffers
sufficient energy to complete any task. \sys's atomicity property derives from
its memory model and data privatization mechanism.

\subsection{\sys Memory Model and Data Privatization}
\label{sec:shared-var}

\sys's memory model provides a familiar programming interface
allowing tasks to share data via global variables. At the same time,
the memory model design allows an efficient implementation of the
task-atomicity guarantee.  The \sys memory access model divides data into {\em task-shared} and {\em
task-local} data.  Multiple tasks or multiple different executions of the same
task may share data using task-shared variables. Task-shared variables are
named in the global scope and are allocated in non-volatile memory.
Task-shared variables have a typical load/store interface: once a task wrote
a value to a task-shared variable, that same task or another task may later read the
value by referencing the variable name. Task-local variables are scoped only to a
single task, must be initialized by that task, and are allocated in the
efficient volatile memory.  

As discussed in Section~\ref{sec:background:consistency}, directly manipulating
non-volatile memory in an intermittent execution can leave data inconsistent
due to \war dependencies.  To prevent these inconsistencies, Alpaca {\em
privatizes} task-shared variables to a task during compilation.  Privatization
creates a task-local copy of a task-shared variable in a {\em privatization
buffer}.  As the task executes, it manipulates the copies in the privatization
buffer. When the task completes it copies data to a {\em commit list} that the
task uses to atomically commit all updates buffered in the privatization
buffer. Section~\ref{sec:privatize} describes how privatization works and why
it is sufficient to keep data consistent.  We emphasize, however, that from the
programmer's perspective, privatization is invisible. To support our
privatization analysis, the programmer need only specify (1) tasks and (2)
task-shared variables. With this information alone, \sys provides its
consistency guarantee automatically and efficiently.

The new syntatic elements \sys introduce is summarized in Table~\ref{tbl:syntax}.

\begin{table}[h]
\small
    \caption{\label{tbl:syntax} {Summary of \sys keywords.} }
	\begin{tabular}{ l | l }
		{\bf Keyword} & {\bf Description} \\ \hline
		{{\tt task}} & {Identifies a function as an \sys task.} \\
		{{\tt \trans}} & {Ends a task and start another task.} \\
		{{\tt \taskshared}} & {Identifies a variable as task-shared.} \\
		{{\tt \entrytask}} & {Task that executes when the device boots for the first time.} \\
		{{\tt \initfunc}} & {Function that executes on every reboot, to reinitialize peripherals.} \\ \hline
    \end{tabular}
\end{table}

\section{\sys Implementation}\label{sec:impl}

Our prototype implements the programming
model defined in Section~\ref{sec:alpacalang} using a compiler analysis and
a runtime library.  The key requirements for an \sys implementation are (1) preserving progress at the
granularity of tasks, (2) ensuring that task-shared and task-local data are
consistent, and (3) doing so efficiently.

To meet these requirements, our \sys implementation uses two techniques.
The first technique is {\em data privatization}, which ensures that data remain
consistent by transparently copying selected values into temporary buffers and
redirecting the task's accesses to the buffer.
The second technique is {\em two-phase
commit}, which both preserves progress and guarantees that a
completed task's updates to its privatized values are all rendered
consistently in memory. \sys's use of {\em task-based execution} is the
foundation of its efficient support for privatization and two-phase commit.

\subsection{Task-Based Execution}
\label{sec:task-based-exec}

\sys tasks are void functions with arbitrary code identified with
the \task keyword.  \sys maintains a global {\tt cur\_task} pointer
in non-volatile memory that records the address of the task
that began executing at the last successful task transition.
\sys also maintains a global non-volatile 16-bit counter, \numBoots, which is
initially 1, is incremented on each reboot or task transition, and is reset to
1 when it reaches its maximum value.  The counter is used to privatize arrays
efficiently (Section~\ref{sec:privatize:array}).  To transition from one task
to the next at a \trans statement, \sys assigns {\tt cur\_task} to the address
of the next task and jumps to the start of that task.  When task execution
resumes after a power failure, control transfers to the start of {\tt
cur\_task}.

\subsection{Privatization}
\label{sec:privatize}

\sys privatizes a subset of task-shared variables in a task to keep them
consistent in case power fails as the task executes. We describe privatization
of scalar (i.e., non-array) data first. Privatization of arrays is described
later in Section~\ref{sec:privatize:array}.  To privatize a variable, \sys
statically allocates a {\em privatization buffer} and
copies the variable that may become inconsistent to its local privatization buffer.
\sys re-writes subsequent memory access instructions to refer to the
privatization buffer instead of the original memory location of
the variable.  At the end of the task, right before the transition to the
following task, \sys commits any changes made
to the privatized copy to its original location, using the \emph{two-phase
commit} procedure (Section~\ref{sec:twopc}). Privatization ensures that tasks execute
idempotently because updates to memory are committed only after
a task has completed.  Idempotent execution ensures that a task's effects are
atomic, which is one of \sys's main language-level guarantees.

The correctness and efficiency of \sys's privatization analysis rely on
several key properties of \sys's design.  For efficiency, \sys does not
privatize all task-shared variables.  Instead, \sys detects \war dependencies
during compilation and privatizes only the variables involved in the
dependencies (Section~\ref{sec:background}).  To identify affected variables, \sys
performs an inter-procedural, backward traversal of each task's control-flow graph,
tracking accesses to each task-shared variable along each path.  If a write and
then a read to the same task-shared variable are encountered along {\em any
path} in the backward traversal, \sys privatizes that task-shared variable.

\sys's compiler generates the instructions for privatizing a variable.  The compiler
first allocates a privatization buffer in non-volatile memory for each variable
that needs to be privatized.  At the beginning of the task, the compiler
inserts code that copies the variable value from its original location to its
privatization buffer. Then, the compiler replaces each reference to the original value
inside the task with a reference to the privatization buffer.
Before each \trans statement, the compiler invokes the first phase of the two-phase
commit operation, \precommitfunc (Section~\ref{sec:twopc}), passing as
arguments the addresses of the original variable and its privatization buffer
along with its size.
%
\begin{wrapfigure}{t!}{0.5\textwidth}
	\centering
		\includegraphics[width=0.49\textwidth]{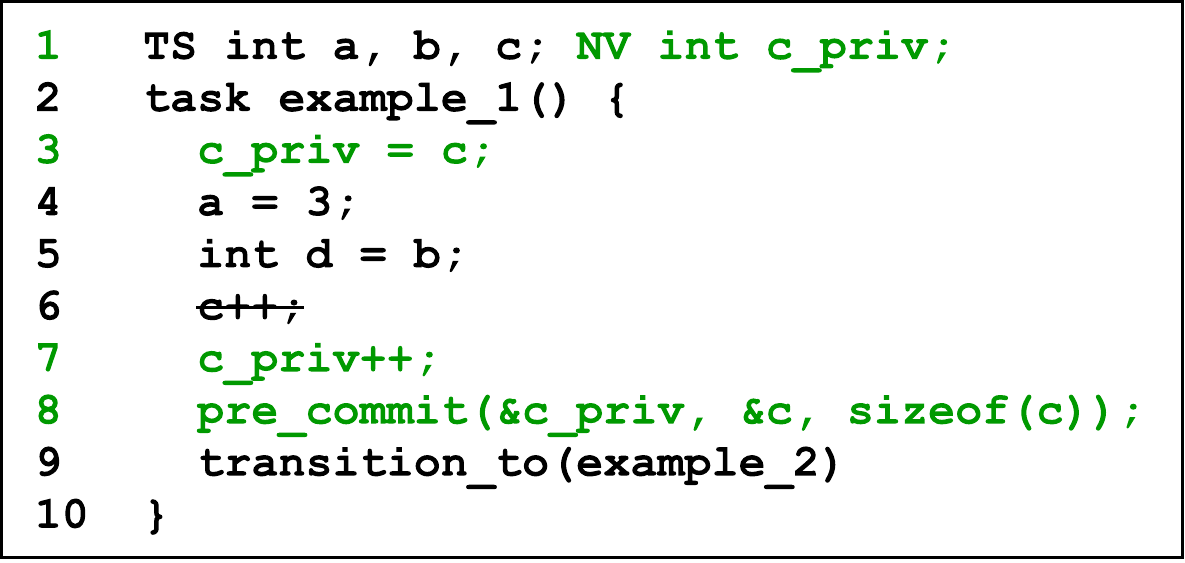}
	\caption{{Privatization and commit.} \trans calls {\tt commit}.}
	\label{fig:a-priv-na}
\end{wrapfigure}

Figure~\ref{fig:a-priv-na} shows a sketch of \sys's instrumentation for an
example task code.  Compiler-inserted privatization code is in green and code
deleted by the compiler is struck-through. As in Line 1, the user defines task-shared variable by annotating
it as {\tt TS}. {\tt TS} variables are saved in non-volatile memory. The code in this example requires
only {\tt c} to be privatized because it is the only 
\war variable; code accessing all other data requires no instrumentation.
Variable {\tt c} is privatized on Line 3, and the access to it on
Line 6 is re-written to refer to the private copy {\tt c\_priv}
(Line 7).  After privatization, only the commit operation can
modify the location {\tt c}.   \emph{Selective} instrumentation
avoids runtime overhead and is the key to \sys's high performance.

Our implementation of the compiler analysis privatizes variables in functions
called from multiple tasks, assuming the variable requires privatization in 
some of its callers. During analysis of a task that calls a function that accesses
such a variable, the compiler rewrites the function's body to refer to
the variable's privatized copy. Consequently, the variable will remain
privatized for any other caller of the same function, even if that caller does
not involve the variable in a \war dependency. This ``contagious''
privatization is safe, conservative, and could be eliminated by replicating the
function body, creating a version for each combination of privatized and
non-privatized variables that the function refers to.  We allow contagious
privatization in favor of the code bloat from replication. In practice,
redundant privatization is rare in the benchmarks that we studied. 

Algorithms~\ref{alg:main}---\ref{alg:transform} depict \sys's privatization
analysis.  The analysis identifies variables potentially involved in \war
dependences, adds code to privatize those variables, and adds code to
atomically commit privatized copies when a task completes.  The code at the end
of Algorithm~\ref{alg:main} identifies the largest possible number of
variables that may need to be committed by a single task and statically
allocates a commit list that accommodates them all. Section~\ref{sec:twopc}
explains in detail how \sys uses its \dirtylist to commit privatized data.

\begin{algorithm}
\begin{algorithmic}[1]
\Function{AlpacaCompiler}{Module $M$}
\For{$t \in M.tasks$} 
	\State $warSet \gets $ \Call{AlpacaFindWAR}{$t$} \Comment{Find \war variables}
	\State \Call{AlpacaTransform}{$t$, $warSet$} \Comment{Modify code for \war variables}
	\State $maxCommitListSize \gets $ \Call{Max}{$maxCommitListSize$, $warSet.size$}
\EndFor
\State \Call{SetCommitListSize}{$maxCommitListSize$} \Comment{Determine \dirtylist size}
\EndFunction
\end{algorithmic}
\caption{Pseudo-code for \sys Compiler.}
\label{alg:main}
\end{algorithm}

\begin{algorithm}
\begin{algorithmic}[1]
\Function{AlpacaFindWAR}{Task $t$}
\State $warSet \gets \emptyset$
\For{$i \in $ $t.instructions$}
	\For{$v \in  i.possibleWriteAddress$} \Comment{Find writes}
		\If{$v \in taskSharedVariables$} 
			\State $i.writeSet \gets i.writeSet  \cup v$
		\EndIf
	\EndFor
	\For{$v \in i.possibleReadAddress$} \Comment{Find reads}
		\If{$v \in taskSharedVariables$} 
			\State $i.readSet \gets i.readSet  \cup v$
		\EndIf
	\EndFor
\EndFor
\For{$i \in $ $t.instructions$} \Comment{Detect \war}
	\For{$j \in i.possiblePreviousInst$} 
		\For{$v \in i.writeSet \cap j.readSet$} 
			\State $warSet \gets warSet  \cup v$
		\EndFor
	\EndFor
	\If{$i.isFunctionCall$} \Comment{For function call (See Section~\ref{sec:privatize})}
		\State $f \gets i.getCalledFunction$
		\For{$v \in f.usedTaskSharedVariables$}
			\State $warSet \gets warSet  \cup v$
		\EndFor
	\EndIf
\EndFor
\State \Return $warSet$
\EndFunction
\end{algorithmic}
\caption{Function Finding \war Variables for Each Tasks.}
\label{alg:analysis}
\end{algorithm}

\begin{algorithm}
\begin{algorithmic}[1]
\Function{AlpacaTransform}{Task $t$, Set $warSet$}
\For{$v \in $ $warSet$}
	\If{$v.isPrivatizationBufferAbsent$} \Comment{Create privatization buffer}
		\State \Call{CreateBuffer}{$v$}
	\EndIf
	\State \Call{InsertPrivatizationCode}{$t$, $v$} \Comment{Insert privatization code}
	\For{$i \in t.instructions$}
		\If{$v \in i.usedOperands$} \Comment{Redirect accesses}
			\State \Call{RedirectUsageToBuffer}{$i$, $v$}
		\EndIf
		\If{$i.isTransitionTo$} \Comment{Insert \precommit code}
			\State \Call{InsertPrecommitBefore}{$i$, $v$}
		\EndIf
	\EndFor
\EndFor
\EndFunction
\end{algorithmic}
\caption{Function Inserting Privatization and Pre-commit Code When Needed.}
\label{alg:transform}
\end{algorithm}

\subsection{Committing Privatized Data}
\label{sec:twopc}

\begin{figure*}
	\centering
	\begin{subfigure}[b]{0.3\textwidth}
		\includegraphics[width=\textwidth]{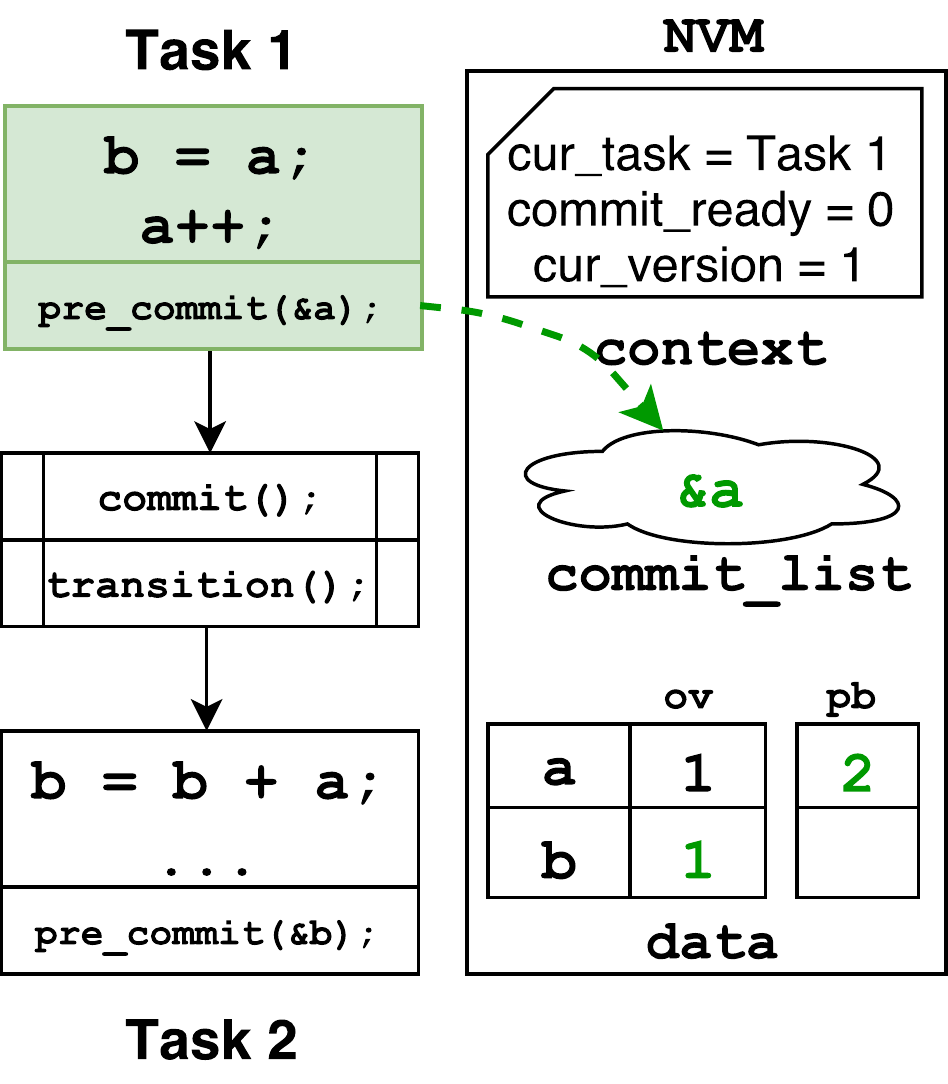}
		\caption{{Executing task 1}}
		\label{fig:flow0}
	\end{subfigure}\hfill%
	\vspace{10pt}
	\begin{subfigure}[b]{0.3\textwidth}
		\includegraphics[width=\textwidth]{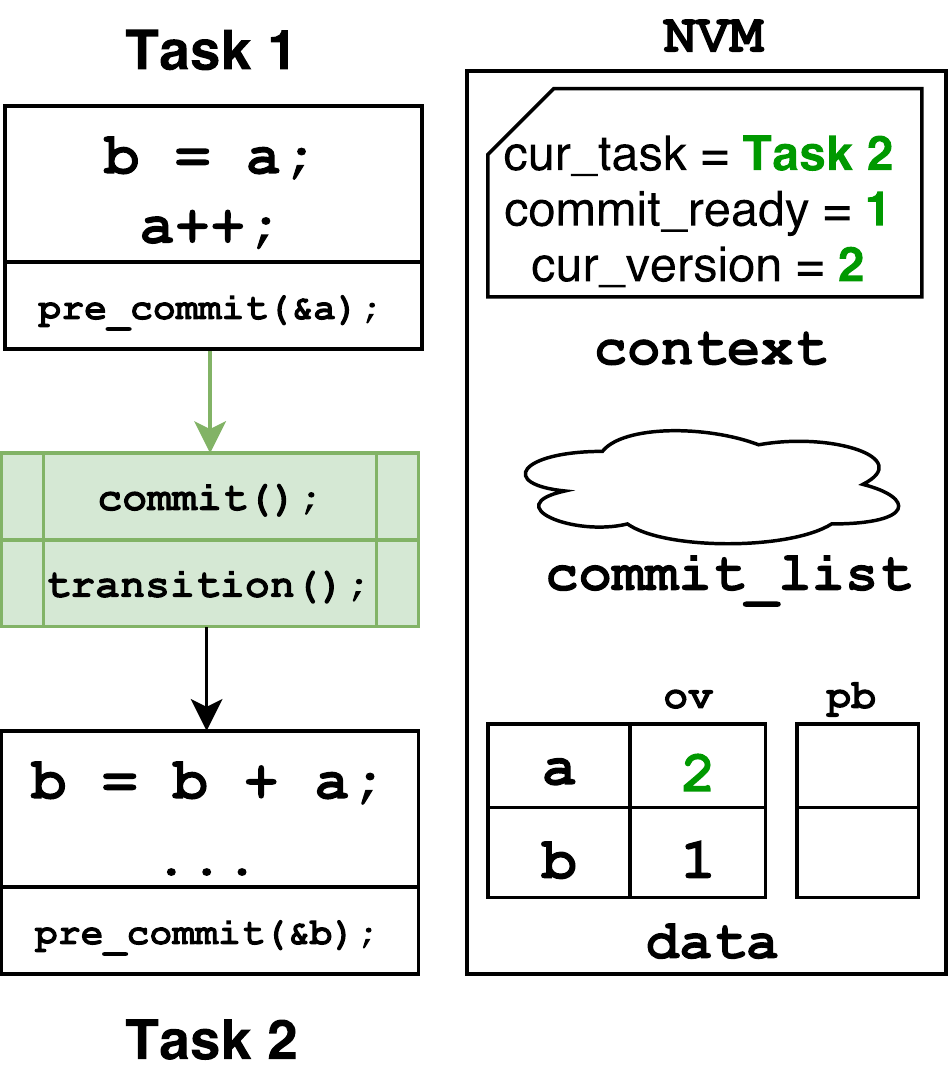}
		\caption{{Committing task 1}}
		\label{fig:flow1}
	\end{subfigure}\hfill%
	\vspace{10pt}
	\begin{subfigure}[b]{0.3\textwidth}
		\includegraphics[width=\textwidth]{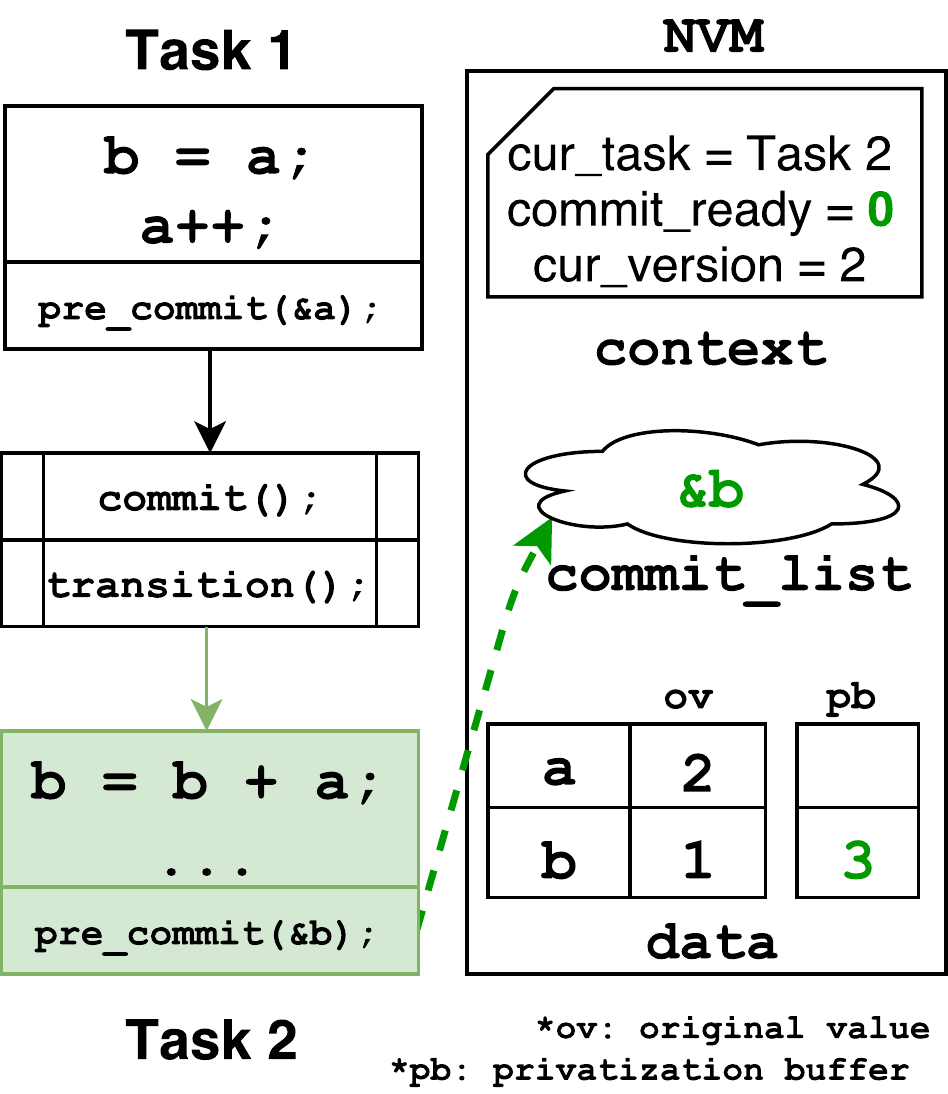}
		\caption{{Executing task 2}}
		\label{fig:flow2}
	\end{subfigure}
	\caption{{Making progress in \sys.} Each panel shows the execution
at left and the system state at right.  The current phase is shaded.  We omit
privatization instructions for clarity. The system state shows that {\tt a} in
Task 1 and {\tt b} in Task 2 are privatized into privatization buffers marked
{\tt pb}.  Initially, {\tt a=1} and {\tt b=0}.  (a) Task 1 writes to {\tt b}
directly and writes to {\tt a}'s privatization buffer because {\tt a} is
involved in a \war dependence.  Updates to privatized variables are written to
the \dirtylist during the {\tt \precommit} phase of the task. A power failure
during execution or {\tt pre\_commit} restarts at the beginning of the task.
(b) Task 1 proceeds to the {\tt commit} phases where Task 1 applies its update
to {\tt a}. A power failure during commit restarts in commit.  (c) The \trans
operation atomically begins Task 2, which privatizes {\tt b} because Task 2
reads then writes it.}
	\label{fig:flow}
\end{figure*}

At the end of a task's execution (i.e., upon reaching a \trans statement)
Alpaca performs a two-phase commit of updates made to privatized data by that
task.  The commit operation atomically applies all updates to variables' original
locations.  The operation is divided into two phases: \precommit and \commit.
The \precommit operation is implemented by
the \precommitfunc  function in \sys runtime library.  This function takes
the variable information as an argument and records it in an entry in
the \dirtylist table, depicted in Figure~\ref{fig:flow0}. The \dirtylist is a
table with exactly one entry for each privatized variable.  A variable's
\dirtylist entry contains the variable's original address, privatization
buffer's address, and size.  Calls to {\tt pre\_commit} are inserted by the
compiler at \trans statements, as was described in
Section~\ref{sec:privatize}. 

The \dirtylist generated in the first phase records updates to privatized data
that must be committed in the second phase.
\sys stores an \commitendindex  that always points to the entry after
the last valid entry in the \dirtylist.  The \dirtylist must be stored in
non-volatile memory since its contents must persist if a failure happens
during the second phase. As seen in Algorithm~\ref{alg:main}, 
our implementation statically allocates a region of
memory large enough to fit the maximum number of entries that may be
required by any task in the program (i.e., the maximum number of calls to
\precommitfunc at any \trans statement in any task). After the last
\precommitfunc call before each \trans, the compiler inserts an instruction
to set a non-volatile \commitready bit that marks the task ready for the second
phase, as shown in Figure~\ref{fig:flow1}.  \sys runtime checks \commitready on
boot. If \commitready is unset, the previously executing task was either in
progress or had completed only a partial \precommit, so that task is
re-executed from its start, discarding the partial execution or the
partial \precommit.  Otherwise, the second phase is invoked.

The second phase, \commit, is implemented in the \sys runtime library by a void
function, \commitfunc.  The function iterates over entries in the \dirtylist
from the first up to \commitendindex. For each entry, the variable value is copied
from its privatization buffer to its original memory location.
The \commit operation succeeds when it copies all entries in the \dirtylist and
sets \commitendindex to zero.
After a successful commit, the runtime clears the \commitready bit and proceeds to the following task
(Figure~\ref{fig:flow2}).  If power fails during \commit, \commitready remains
set.  Since the runtime checks the bit on boot, it will retry the \commit operation
until it completes successfully.  If power fails after \commit but before
\trans completes the transition to the next task, then \commit will re-execute on
next boot and will trivially complete since \commitendindex is zero. The \trans that
failed to complete will then run again.
\vspace{10pt}

\subsection{Privatizing and Committing Arrays}
\label{sec:privatize:array}
\begin{wrapfigure}{R}{0.46\textwidth}
	\centering
		\includegraphics[width=0.45\textwidth]{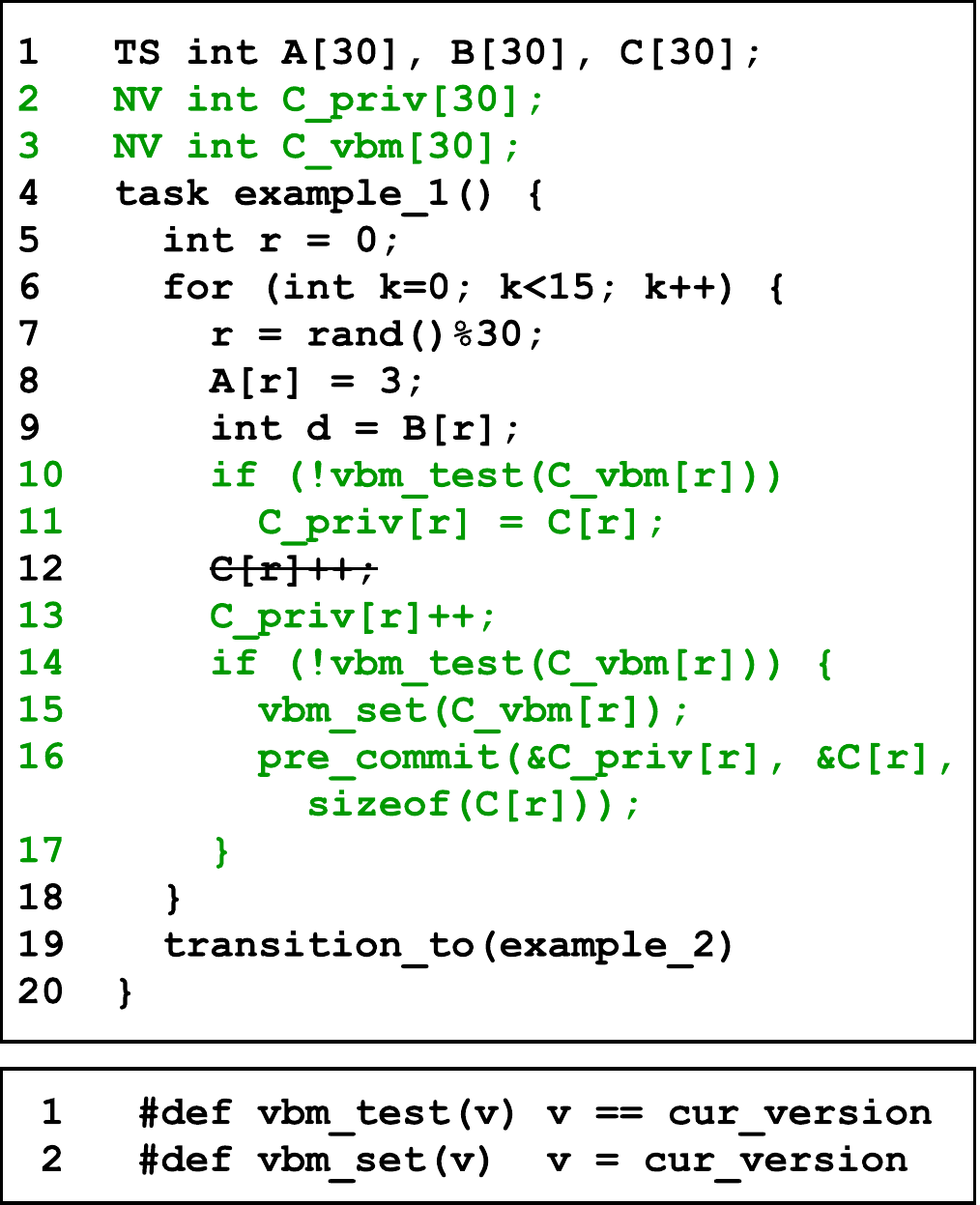}
	\caption{{Privatization and commit for arrays.} }
	\label{fig:a-priv-a}
\end{wrapfigure}

\sys privatizes and commits array variables differently from scalar variables
because naively privatizing an entire array (i.e., copying the entire array to
a privatization buffer as a task starts) is unnecessary if the task accesses
only part of the array.  \sys statically pre-allocates a privatization buffer
for each array that may be read then written (i.e., may be involved in a \war
dependence). The array's privatization buffer contains the same number of
entries as the original array. Privatization takes place at the granularity of
an array element.  In the example in Figure~\ref{fig:a-priv-a}, to privatize
array {\tt C}, the compiler allocates {\tt C\_priv} buffer (Line 2) and inserts
the instrumentation code that is highlighted in green (and explained below).

Like a scalar variable, privatizing an array element involves initializing a
copy in the privatization buffer (Line 11), redirecting accesses to the buffer
(Lines 12-13), and adding the variable to the \dirtylist via a call to
\precommitfunc (Line 16).  \sys uses the compiler to redirect array element
accesses to their privatization buffers the same as for scalars, but
initializing privatization buffers and pre-commit for arrays are different.
\sys initializes an array element's privatization buffer the first time an
execution accesses the element: either explicit instrumentation inserted by
\sys initializes the buffer before the element's first read {\em or} the
element's first write directly writes to the buffer.  
\sys does pre-commit for an array element only once after the first write to
it.  

One key design choice in \sys was to decide when instrumentation on a read
operation should initialize an array element's privatization buffer.  Read
instrumentation should not initialize the privatization buffer after a previous
write in the task because the initialization would overwrite the written value.
Instead, the read instrumentation can initialize the privatization buffer
either once before the first read that happens before the first write or
(possibly redundantly) at every read before the first write.
We chose the latter option to avoid the overhead of dynamically tracking the
first read, which incurs a high runtime overhead. 

We avoid invoking pre-commit unconditionally after every write because multiple
writes to the same element would append duplicate entries to the \dirtylist,
which is inefficient and precludes a statically sized \dirtylist.
Furthermore, pre-commits cannot be batched and executed before a task
transition (like for scalar variables), because the set of elements dynamically
accessed is unknown statically. Batching would require dynamically tracking the
set of modified elements in a data structure that supports efficient insertion
and traversal which is complex.
Executing pre-commit after the first write obviates the complexity of batching
and only requires \sys to identify the first write to an array element.

Correctly handling array privatization and pre-commit requires some
instrumentation to execute conditionally, only on an element's first write.
To identify an element's first write, \sys must track the set $U$ of array
elements that have been written since the beginning of the task in the current
execution attempt.
A write of an element is first if and only if the element is not in this set
$U$ at the time of the access.
The data structure that represents $U$ needs only to provide efficient
insertion and lookup, which our \emph{version-backed bitmask} data structure
does.
A version-backed bitmask is a bitmask that supports a constant-time clear
operation using a versioning mechanism described later in this section.
We represent $U$ by setting logical bits (i.e., ``entries'') in a
version-backed bitmask that is statically allocated for each array being
privatized.
In Figure~\ref{fig:a-priv-a}, the version-backed bitmask for {\tt C} is {\tt
C\_vbm} allocated on Line 3.

Each version-backed bitmask entry is a 16-bit integer \emph{version}. To set an
entry ({\tt vbm\_set}), \sys copies the global \numBoots counter into the
entry. To test an entry ({\tt vbm\_test}), \sys compares the version stored in
that entry to the global \numBoots counter; equality indicates the entry is
set, inequality indicates unset.  Consequently, when the global \numBoots
counter changes, all version-backed bitmasks are implicitly cleared. When the
\numBoots counter overflows and rolls over, the runtime explicitly resets all
entries in all version-backed bitmasks to zero.

To track the set $U$ of array elements updated in the current task execution
attempt, the \sys compiler instruments reads and writes to array elements with
code to set and test entries in the array's version-backed bitmask.
When reading from an array element that has not been modified yet, i.e. its
entry in $U$ is unset (Line 10), then the runtime initializes the element copy
in the privatization array (Line 11). When writing to an array element for the
first time, after checking that its entry in $U$ is not set (Line 14), it
inserts the element into $U$ by setting its entry (Line 15), and appends the
written array element to the \dirtylist by calling \precommitfunc (Line 16).
The set $U$ is cleared at the next task transition or reboot, since the \numBoots
counter increments on each task transition and reboot
(Section~\ref{sec:task-based-exec}), which implicitly clears the version-backed
bitmask.


\section{\sys with Undo-Logging}\label{sec:undo}

The design of \sys described up to this point relies on privatization and
commit, which is a {\em redo-logging} approach~\citet{alpaca} to keeping memory
consistent across intermittent failures.  
We also developed a more efficient \sys design variant that relies instead on
undo-logging.
Both design variants use the same programming interface, differing only in how
they manage memory.
Section~\ref{sec:eval} compares undo- and redo-logging, showing that
Alpaca-undo is on average 1.53x faster than its redo-logging counterpart.
This text refers to the \sys design using privatization and commit as {\em
\sys-redo}, refers to the undo-logging variant as {\em \sys-undo}, and refers
to both generally as \sys. 

\subsection{Undo-Logging Instead of Privatization}
\label{sec:undo-vs-redo}

Redo-logging and undo-logging each present advantages and disadvantages.
\sys-redo privatizes variables involved in \war dependences and commits updates
to those data, when a task completes. 
Redo-logging affords {\em zero-cost recovery}: requiring no action before
continuing after a power failure. However, redo-logging pays a cost in its need
to first privatize data and later commit them, which requires {\em two} copy
operations per variable per completed task.
In contrast, undo-logging backs up a variable and subsequently manipulates the
variable in place, requiring no action when a task completes.  However, an
undo-logging system must restore values saved in the undo log before continuing
execution after a power failure. While restoring from power failure is more
costly than in a redo-logging system, undo-logging requires only {\em one} copy
operation (to back a variable up) per variable per completed task.
When successful completion of a task is more common than the interruption of a
task by a power failure, undo-logging will be more efficient than redo-logging.
In \sys, successful task completion is usually more common than interruption by
a power failure because a typical task requires much less energy than the maximum
energy that the device can buffer.
In the worst case when all tasks require the maximum amount of energy that the
device can buffer, the task will fail once for each completion.
Under \sys's assumptions, undo-logging is appealing because tasks cannot fail
more often than they complete.

\subsection{Undo-logging Compiler and Runtime System}
\label{sec:undo-comp}
\label{sec:undo-runtime}

Similarly to \sys-redo, \sys-undo relies on a compiler to transform code and
insert calls to the \sys-undo runtime system into the program.

\begin{wrapfigure}{R}{0.42\textwidth}
	\centering
		\includegraphics[width=0.41\textwidth]{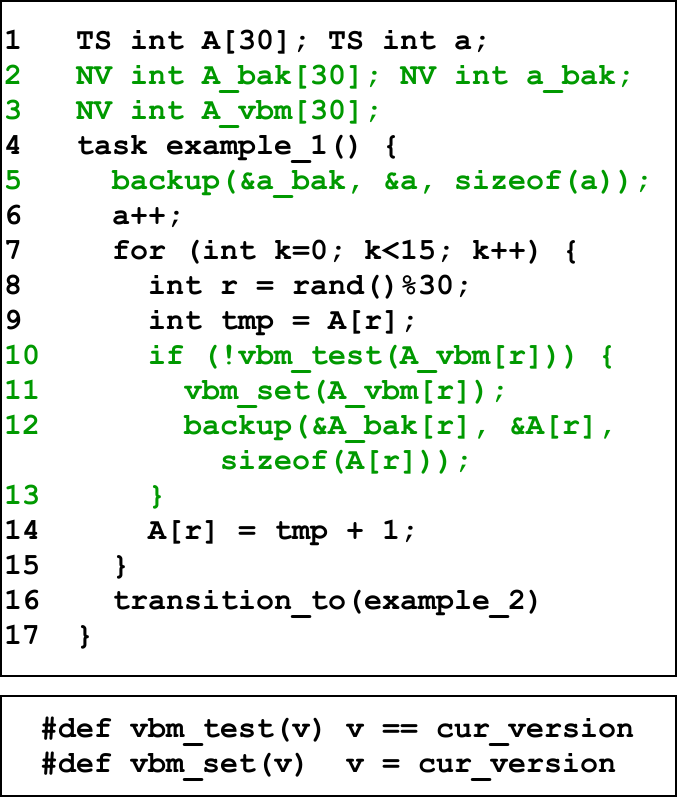}
	\caption{{\sys-undo's compiler transformation.} }
	\label{fig:a-priv-undo}
\end{wrapfigure}

Figure~\ref{fig:a-priv-undo} shows how \sys-undo transforms code to implement
undo-logging.  Instead of allocating a private copy, the compiler allocates a
static undo log (Line 2).  \sys-undo selectively backs up non-array \war
variables at the start of the task (Line 5).  For an arrays, \sys-undo uses
\sys's version-backed bitmask scheme to detect the first write, as discussed in
Section~\ref{sec:privatize:array} (Line 10). \sys-undo backs up an array value
before its first write (Line 12).  Unlike \sys-redo, \sys-undo need not
redirect memory accesses to a copy because operations manipulate data in place
(Line 6, 9, 14).  Additionally, \sys-undo does not need instrumentation before
an array read (Line 9).
\sys-undo detects variables involved in \war dependences using
Algorithms~\ref{alg:main}---\ref{alg:transform}, identically to \sys-redo.

\begin{figure*}
	\centering
	\begin{subfigure}[b]{0.3\textwidth}
		\includegraphics[width=\textwidth]{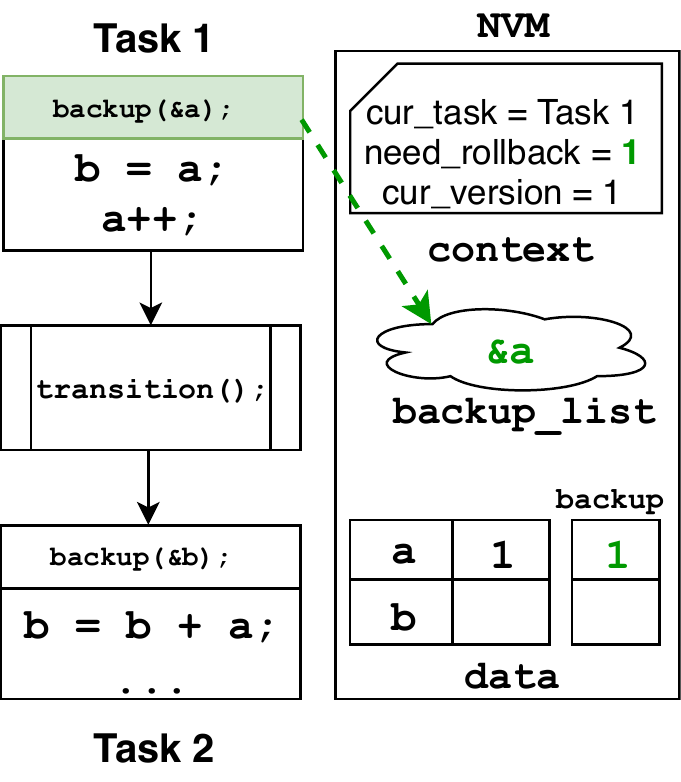}
		\caption{{Backup of variable {\tt a}}}
		\label{fig:flow0_undo}
	\end{subfigure}\hfill%
	\vspace{10pt}
	\begin{subfigure}[b]{0.3\textwidth}
		\includegraphics[width=\textwidth]{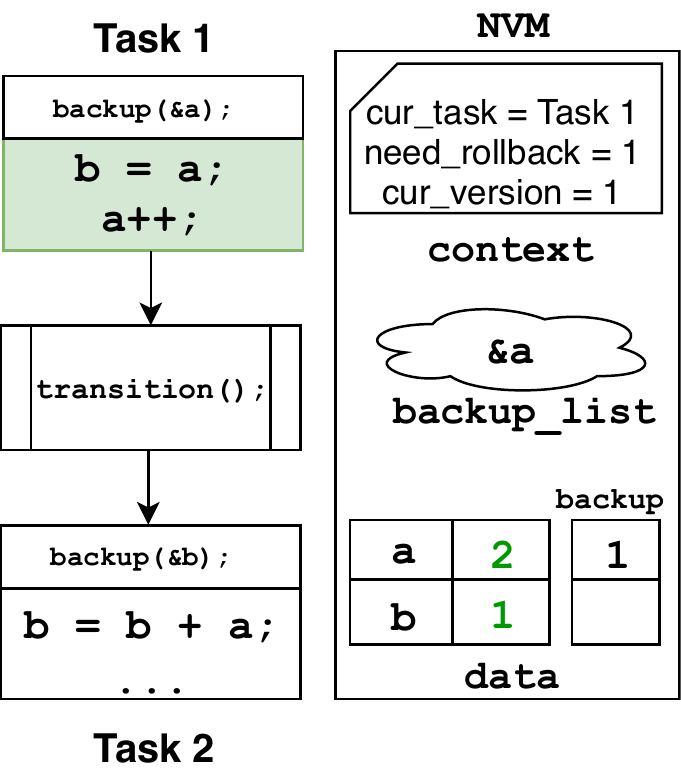}
		\caption{{Executing task 1}}
		\label{fig:flow1_undo}
	\end{subfigure}\hfill%
	\vspace{10pt}
	\begin{subfigure}[b]{0.3\textwidth}
		\includegraphics[width=\textwidth]{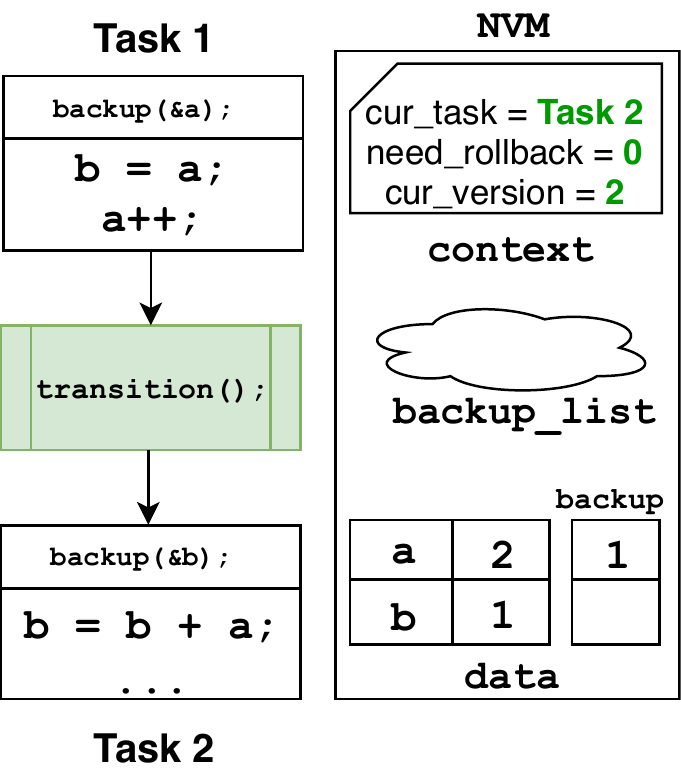}
		\caption{{Finishing task 1}}
		\label{fig:flow2_undo}
	\end{subfigure}
	\caption{{Making progress in \sys-undo.} Each panel shows the execution
at left and the system state at right.  The current phase is shaded.
Initially, {\tt a=1} and {\tt b=0}.  (a) Before executing Task 1, Alpaca-undo
copies the value of {\tt a} to its backup copy. Backed up variables are marked
in {\tt backup\_list}.  (b) Task 1 gets executed, updating variables in-situ.
(c) When Task 1 is finished, Alpaca-undo simply clears the {\tt backup\_list}
and related flags.}
	\label{fig:flow_undo}
\end{figure*}

Figure~\ref{fig:flow_undo} shows how \sys-undo backs up and restores data to
keep memory consistent.  At the beginning of a task, \sys-undo backs up the
task's \war variables (Figure~\ref{fig:flow0_undo}).  \sys-undo maintains a
list of backed-up variables ({\tt backup\_list}).  \sys-undo also sets the {\tt
need\_rollback} flag, indicating that there are variables backed up. Then,
Alpaca-undo manipulates variables in place (Figure~\ref{fig:flow1_undo}).  Even
after the update to {\tt a} its original value remains in the backed-up copy.
When a task successfully completes, \sys-undo clears the {\tt need\_rollback}
flag and {\tt backup\_list} (Figure~\ref{fig:flow2_undo}). The runtime system
clears the list efficiently by resetting the list's iterator, without zeroing
its contents.  After a power failure, \sys-undo rolls back changes by iterating
through the {\tt backup\_list} if the {\tt need\_rollback} flag is set. For
each entry, the runtime system writes the backed-up value into its
corresponding memory location. After processing all entries, the runtime unsets
the {\tt need\_rollback} flag and continues.
Section~\ref{sec:privatize} discusses sizing the backup list. 

\section{\sys Discussion}
\label{sec:discussion}

\sys's programming model guarantees that tasks will execute atomically. Our
\sys implementation efficiently provides this atomicity guarantee by
selectively privatizing data.  Besides programmability, efficiency, and
consistency, \sys  supports I/O operations and allows modular re-use of code.
This section discusses these aspects of \sys and characterizes its main
limitations. 

\subsection{Low Overhead}

A key contribution of this work is that our \sys implementation has
low overhead compared to existing systems to which we can directly compare
(we quantify the difference in Section~\ref{sec:eval}).  \sys's overhead
is low, because privatization is simple and because \sys privatizes
variables selectively.   Privatization has a low cost, primarily because it rarely
occurs: most variables are not privatized because they are either local to
a task or shared but not involved in \war dependences.  Furthermore, \sys's
task-based execution avoids all checkpointing cost. \sys needs to retain
only the information about which task was last executing. \sys does not incur
the cost of tens of writes to non-volatile memory to save registers,
like Ratchet~\cite{ratchet}, nor the even higher additional cost to
save the stack, like DINO~\cite{dino}.  By reducing copying and privatizing
only when necessary, \sys saves time and energy.

\subsection{Memory Consistency}

\sys preserves memory consistency despite arbitrarily-timed power failures
by making each attempt to execute a task idempotent.   Task idempotence
guarantees that if any attempt has sufficient energy to complete, the effects
of a single, atomic execution of the task are made visible in memory. The
memory state immediately after a task transition is equivalent
to the corresponding state in execution on continuous power.
\sys guarantees idempotence by privatizing non-volatile variables involved in \war dependences
and requiring volatile state to be task-local.

\subsubsection{Non-volatile Memory Consistency.} 

Taking a cue from prior work~\cite{idempotent, idempotent2, ratchet, dino},
\sys privatizes only non-volatile variables involved in \war dependencies.
%
%
%
We show that privatizing only this subset is sufficient by proving that only
memory accesses related by \war can cause a value written by the
task before a power failure to be read by the same task after the power
failure.

%
Consider one task and assume that control flows along
the same path each time the task re-executes, which is true of all code that
does not perform I/O operations (we discuss I/O later in this section).  Consider
one memory location and let $R_{i}^{j}$ and $W_{i}^{j}$ respectively denote the
$i^{\text{th}}$ memory read and write to that location during the
$j^{\text{th}}$ attempt to execute the task.
If power fails in attempt $j$ after $k$ accesses and the task re-executes, then
the sequence of memory accesses is:
$X_{0}^{j},\dots W_{p}^{j}\ldots X_{k}^{j}-[\text{power
failure}]-X_{0}^{j+1}\dots R_{q}^{j+1}\ldots$, where $X$ stands for either
read or write and our hypothesis postulates a write $W_{p}^{j}$ before
the power failure and a read $R_{q}^{j+1}$ that returns the same value.
%
%
%
%
%
The hypothesis implies that $q < p$, otherwise, $W_{p}^{j+1}$ would overwrite
the value written by $W_{p}^{j}$ before $R_{q}^{j+1}$ reads it.
The order $q < p$ implies that $R_{q}^{j}$ precedes $W_{p}^{j}$ in the
task code, which is the definition of a \war dependence.

\subsubsection{Volatile Memory Consistency.} 
In Alpaca, the only volatile data are task-local variables.  Since all local
variables must be initialized before use in a task, local reads after a power
failure will never access uninitialized memory. Since volatile memory clears on
reboot, local reads will never observe a value written before the power failure.

Like prior work~\cite{ratchet}, \sys conservatively assumes that compiler
optimizations cannot introduce memory read or write instructions and \sys
safely interacts with any compiler optimization that adheres to this
assumption.

\begin{wrapfigure}{R}{0.38\textwidth}
	\centering
	\includegraphics[width=0.37\textwidth]{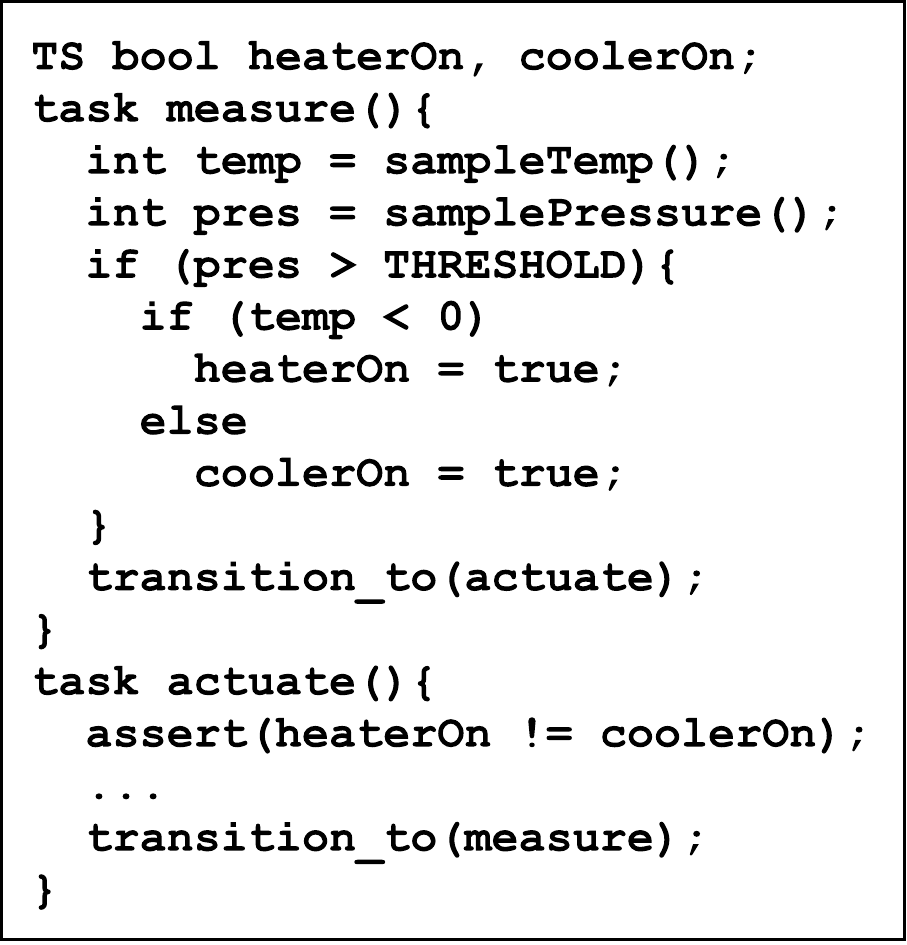}
	\caption{{I/O in \sys.}}
	\label{fig:io}
\end{wrapfigure}

\subsection{I/O}
\label{sec:io}

Code that interacts with sensors and actuators poses three difficulties: (1)
some I/O-related actions must execute atomically, (2) external inputs introduce
non-determinism, and (3) actuation or output cannot be undone. \sys allows the
programmer to express (1) and (2) through careful coding patterns that we
describe below.  \sys targets applications that can tolerate repeated outputs,
where (3) is acceptable.

Some applications include I/O-related code that should execute atomically, such
as the code in Figure~\ref{fig:io}.  The code reads temperature and pressure
sensors and sets the {\tt heaterOn} or {\tt coolerOn} flag, based on the sensed
data.   The temperature and pressure values should be consistent. \sys
lets the programmer ensure that the values will be consistent by putting the
actions in the same task.  In contrast, a system with
dynamic~\cite{mementos,hibernus} or compiler-inserted~\cite{ratchet} task
boundaries gives the programmer no way to ensure that the input operations
execute atomically.

The code in the example asserts that {\tt heaterOn} and {\tt coolerOn} are
never both {\tt true}.  The code misbehaves if a power failure occurs after
assigning one of the flags (e.g., {\tt heaterOn}). If the sensor's result is
different in the following execution attempt, the code could set the other flag
(e.g., {\tt coolerOn}), violating the assertion.  The core issue is that non-volatile
memory updates are conditionally dependent on sensed inputs.  If control-flow
depends on the input, then \emph{conditional} non-volatile memory updates can
violate task idempotence.  We note that this problem also afflicts prior
efforts~\cite{chain,ratchet,dino}.  A programmer can preserve idempotence by using
intermittence-safe I/O programming patterns.
Concretely, one programming pattern that avoids the problem in this example
is to use a dedicated task to read and store both {\tt temp} and {\tt pres}, 
and to use another task to do the conditional updates to {\tt heaterOn} and {\tt coolerOn}. 
Alternatively, a programmer could avoid the problem by ensuring that both
execution paths access the same set of memory locations: inserting {\tt
coolerOn = false;} on the {\tt if} branch and inserting {\tt heaterOn = false;}
on {\tt else} branch.

\subsection{Forward Progress}

Guaranteeing forward progress in an intermittent, energy-harvesting system is a
difficult problem that is orthogonal to the problems solved by \sys. The key
challenge is that a system buffers a fixed amount of energy before it begins
operating and if the energy required by a task exceeds the buffered amount, the
task will never complete executing, preventing progress.  A task's energy cost
can be input dependent, adding further complexity. This progress issue is not
unique to \sys, afflicting prior task-based systems as well~\cite{mementos,
dino, ratchet, chain, idetic}.

Prior work has used {\em ad hoc} techniques that attempt to ensure progress, to
the detriment of other system characteristics.  Ratchet~\cite{ratchet} inserts
a dynamic checkpoint between static checkpoints after repeatedly failing to
make progress.  Other systems~\cite{hibernus,
hibernusplusplus, quickrecall} dynamically checkpoint in response to an
interrupt when energy is low.  Dynamic checkpointing requires capturing enough
state to restart from an arbitrary point, which can take a prohibitive amount
of time~\cite{chain}, especially with hybrid volatile and non-volatile memory.
Dynamic checkpointing may also violate I/O atomicity (see Section~\ref{sec:io}).

We opted not to include a dynamic checkpointing fall-back in \sys. Instead the
programmer must ensure for sizing tasks such that tasks in their
program do not require more energy than their target device can buffer. As
long as this condition is satisfied, \sys always avoids atomicity violations
and guarantees correctness.  None of the tasks in our test programs have a
forward progress problem.  It would
be straightforward to incorporate a dynamic checkpointing fall-back into our
\sys prototype.

\subsection{Reusability of Tasks}


In a task-based programming model for intermittent execution, code reuse
via functions is insufficient, because functionality that 
uses more energy than the device can buffer cannot be encapsulated in a single function.
The programmer of \sys can reuse the sequence of tasks as
a C programmer would reuse a function by
passing arguments, return address, and return value
manually through task-shared variables.

%


\subsection{Prototype Limitations}
\label{sec:limitation} 

Our \sys prototype supports a useful subset of the C language, handling most
uses of pointers and complex data structures.  Our prototype has a few
implementation-specific limitations, which we emphasize are not fundamental
limitations of \sys.  

We implemented a limited pointer alias analysis and our prototype requires that
a {\tt TS} pointer only ever be assigned the address of a {\tt TS} variable if
that address is constant.  Allowing {\tt TS} pointers to constant
variables permits the especially important case of function pointers.
 
Our prototype requires the programmer to refer to array elements directly,
i.e., writing {\tt A[30]} instead of {\tt *(p + 30)}.  Our
prototype statically inserts code to maintain version-backed bitmasks on array
accesses.  Array indirection would require our prototype to use instrumentation
that dynamically disambiguates pointers to arrays, to determine which bitmask
to update.  Our prototype makes the calculated choice to avoid this additional
dynamic analysis cost by requiring direct array access. We note that this
strategy is similar to DINO~\cite{dino}.

\section{Benchmarks and Methodology}
\label{sec:apps}

We evaluated \sys using a collection of applications taken from prior work
running on real, energy-harvesting hardware.  Our evaluation ran on 
a WISP5~\cite{wisp} energy-harvesting platform that runs a TI
MSP430FR5969 microprocessor with harvested RF energy.  
We used a Saleae Digital Logic Analyzer to measure the execution time
by timing GPIO pulses generated at the end of each application.
To power the WISP, we used the ThingMagic Astra-EX RFID
reader as an RF energy source with its power parameter set to {\tt -X 50}, and
a distance between the WISP and the power source of 20cm. 

We evaluated \sys using applications ported to run on harvested energy using
DINO~\cite{dino}, Chain~\cite{chain}, Alpaca-redo, Alpaca-undo, 
and Ratchet~\cite{ratchet}, allowing for a thorough direct
comparison.  DINO and Chain versions of four applications were provided by the
authors of the \chain~\cite{chain} paper: activity recognition (AR), cuckoo
filter (CF), rsa encryption (RSA), and cold-chain equipment monitoring (CEM).
We ported two additional applications from the MIBench~\cite{mibench} to run
with DINO, Chain, and two versions of \sys.

DINO assumes precise pointer aliasing to work correctly~\cite{dino}, and performs
very poorly on conservative, practical pointer aliasing. Thus, the DINO code we
obtained from the author of Chain~\cite{chain} was hand-annotated assuming perfect
pointer aliasing. Our evaluation shows that \sys even outperforms the hand-annotated
{\em oracle} DINO. \sys does not assumes any perfect pointer aliasing as in DINO.

We ported Ratchet~\cite{ratchet}, which was originally targeted for ARM architecture,
to run on TI MSP430 series. While doing so, we lack some of the ARM-specific
optimizations suggested by the original work.
According to the evaluation by the authors,
omitting the optimizations can lead to around 1.6x slowdown~\cite{ratchet}.

We studied six applications, summarized in Table~\ref{tbl:apps}.

\begin{table}[h]
\small
    \caption{\label{tbl:apps} {Summary of benchmarks.} }
	\begin{tabular}{ l | l }
		{\bf App} & {\bf Description} \\ \hline
		{CEM} & {LZW-compresses a random number stream using 512-entry dictionary.} \\
		{CF} & {Stores and retrieve a sequence of random numbers using 128-entry filter.} \\
		{RSA} & {Encrypts a 11-byte string with 64 bit key using RSA encryption.} \\
		{AR} & {Collects 128 accelerometer samples and use nearest neighbor classification to detect movement.} \\
		{BF} & {Encrypts a 32-byte string using Blowfish encryption.} \\
		{BC} & {Counts the number of set bits in a given input stream.} \\ \hline
    \end{tabular}
\end{table}

To ensure a fair comparison, applications use identical task definitions for
Chain, Alpaca-redo, and Alpaca-undo, and we inserted task boundaries at equivalent code points
for DINO.  
Since inserting boundaries automatically is part of the system for Ratchet,
we did not manually inserted boundaries for Ratchet.

\section{Evaluation}
\label{sec:eval}

\begin{figure}[t]
	\centering
	\begin{subfigure}{0.49\textwidth}
		\includegraphics[width=\textwidth]{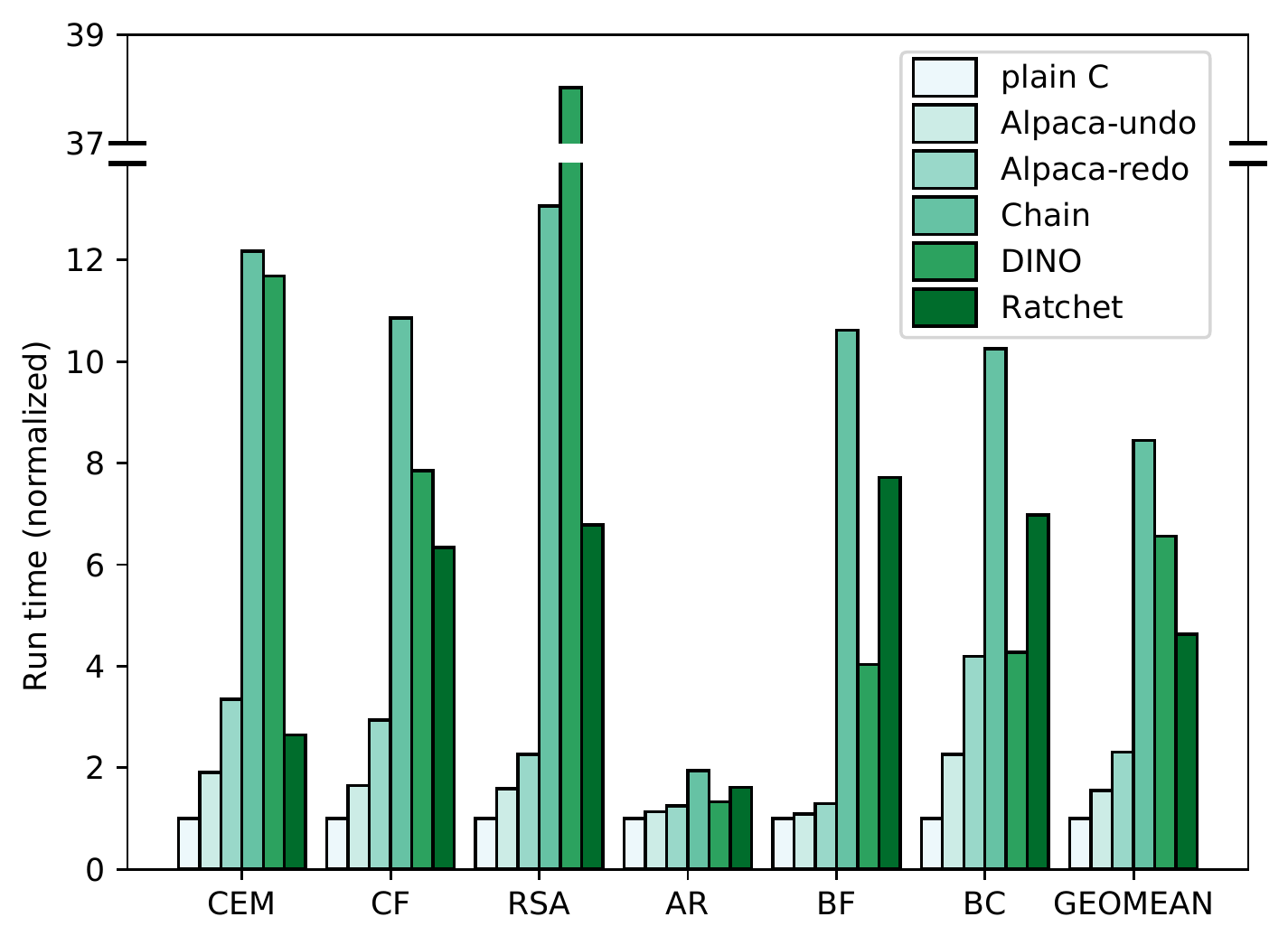}
		\caption{{On continuous power}}
		\label{fig:cont_power}
	\end{subfigure}
	\begin{subfigure}{0.49\textwidth}
		\includegraphics[width=\textwidth]{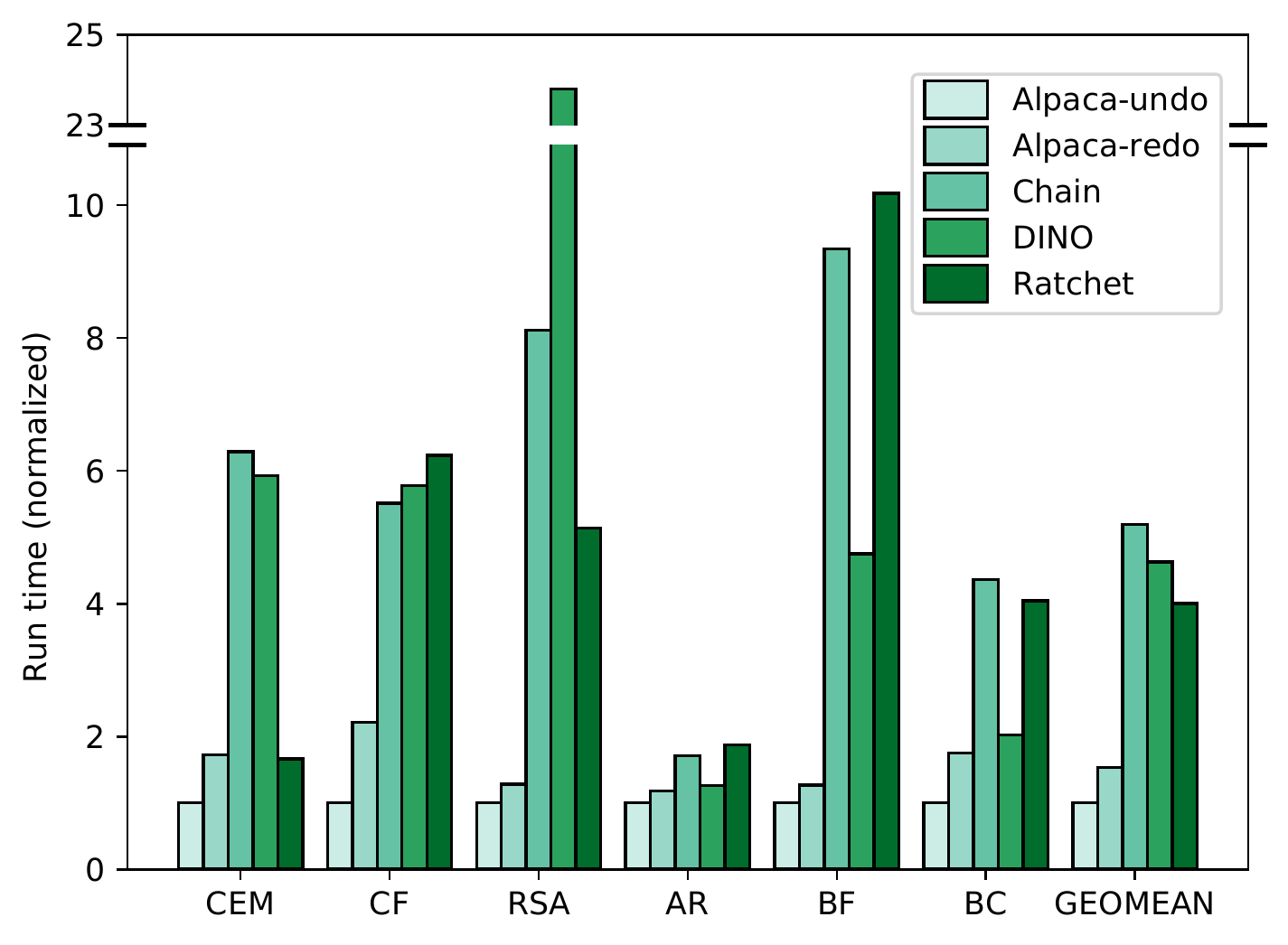}
		\caption{{On harvested energy}}
		\label{fig:int_power}
	\end{subfigure}
	\caption{{Run time performance.} Data are normalized to performance of (a) plain C and (b) Alpaca-undo.}
	\label{fig:runtime}
\end{figure}

Our evaluation compares directly to Chain, DINO, and Ratchet and illustrates several
findings about \sys.  The data show that both Alpaca-undo and Alpaca-redo outperforms existing systems
while running natively on existing hardware both on harvested energy and
running on continuous power.  Our evaluation characterizes these findings,
showing that \sys avoids the costliest time and memory overheads of prior
approaches. We qualitatively and
quantitatively show that programming with \sys is simple compared to other
approaches.  We also contrast our \sys-redo implementation with an alternative \sys-redo
design that privatizes data to volatile memory, showing that our baseline
design is usually more efficient because of additional overheads
required by volatile privatization.

\subsection{Run Time Performance}

\begin{wrapfigure}{t!}{0.46\textwidth}
	\centering
	\includegraphics[width=0.45\textwidth]{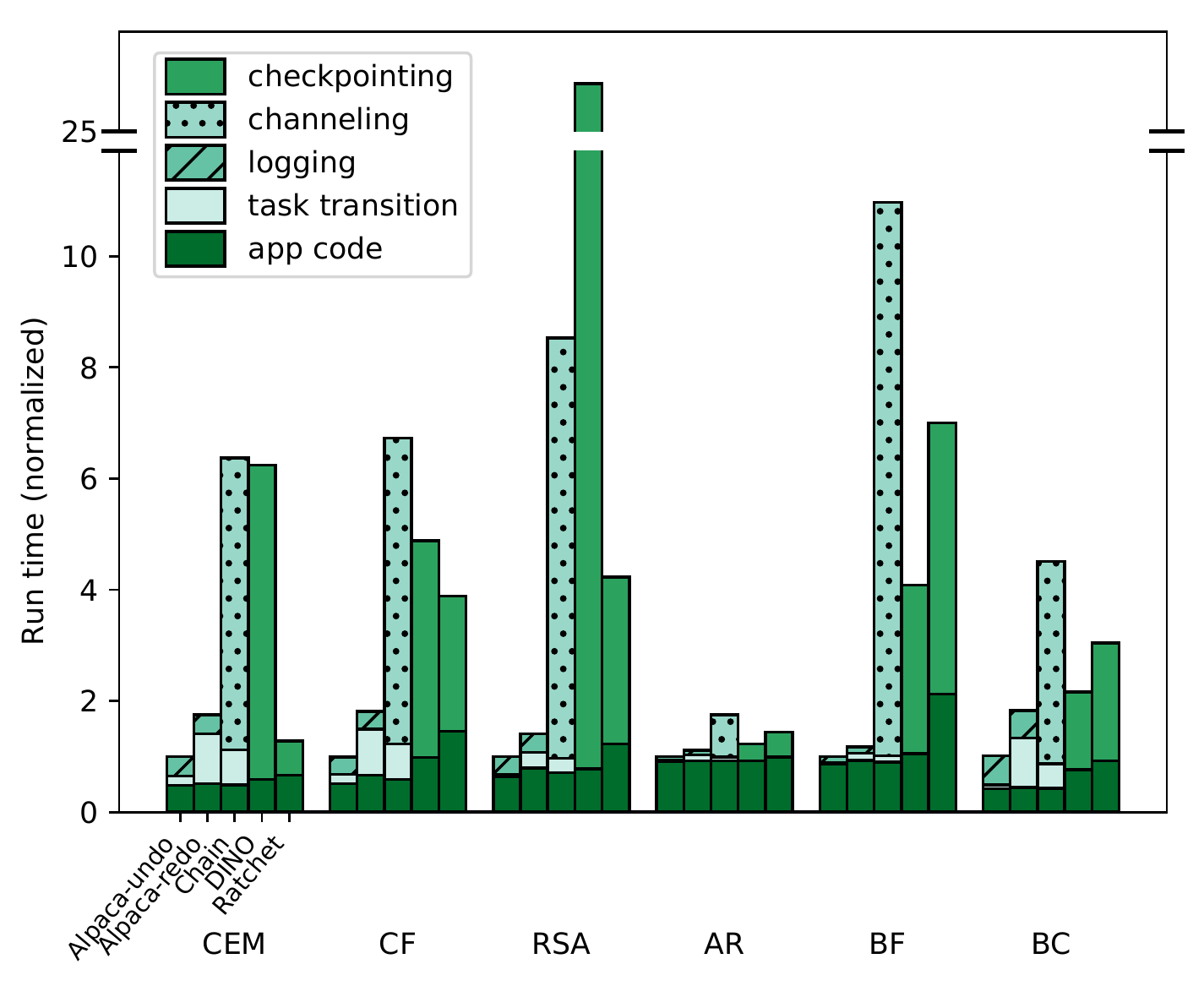}
	\caption{{Breakdown of overheads.} Bars are normalized to the run time of Alpaca-undo.}
	\label{fig:overhead}
\end{wrapfigure}

Figure~\ref{fig:runtime} shows \sys's run time performance, measured {\em on
real hardware} on both continuous power and on harvested RF energy.
Performance on continuous power is an upper bound on performance because it
avoids reboot-related overhead.  Performance on harvested energy includes all
reboot-related overheads and is representative of a real-world deployment.

Figure~\ref{fig:cont_power} shows performance on continuous power for each
system, normalized to a plain C implementation that implements each application
without considering intermittence. 
As expected, \sys has an overhead compared
to plain C code, with the average slowdown of 1.55x (Alpaca-undo) and 2.31x (Alpaca-redo), respectively.  
However, both \sys outperforms previous state-of-the-art systems \chain, \dino, and Ratchet.
Also, Alpaca-undo consistently outperforms Alpaca-redo by 1.49x on average,
exemplifying the optimization discussed in Section~\ref{sec:undo-vs-redo} to be useful.
When compared to the previous state-of-the-arts, Alpaca-undo outperforms \chain, \dino, and Ratchet
by 5.44x, 4.22x, and 2.99x on average.


Figure~\ref{fig:int_power} shows performance on harvested energy. Here, we omit
the plain C variant because it does not handle intermittence and cannot run
correctly on harvested energy.  Alpaca-undo again outperforms Alpaca-redo by 1.53x, and
both \sys outperforms all the other systems, by Alpaca-undo outperforming \chain, \dino, and
Ratchet by 5.19x, 4.63x, and 4.00x on average.
The performance gap is mostly larger on
harvested energy because power failures introduce reboot-related overheads and
Alpaca's reboot overhead is extremely low. 

\subsection{Characterizing \sys's Runtime Overhead}
\label{sec:overhead}

To better understand \sys's performance, we made detailed measurements
of each system's major overheads. The two \sys's main overheads are logging (undo-logging for Alpaca-undo, and
redo-logging or privatization for Alpaca-redo) and
task transitioning. \chain's major overheads are channel manipulation and task
transitioning.  The task transitioning of the three systems are different:
Alpaca-undo's task transitioning simply clears the index of the backup list and some flags,
whereas Alpaca-redo's task transitions commit privatized state and \chain's task
transitions commit all data written to ``self'' channels.  \dino and Ratchet's major
overheads are checkpointing and restoring the checkpoint on reboot.

We measured each system's overheads by toggling GPIO at the beginning and at the end of each overhead and
summing up the duration using Saleae Logic Analyzer. When measuring the overhead, we experimented on
continuous power instead of harvested energy, since frequent GPIO toggling consumes non-negligible amount of energy.
On continuous power, we used the microcontroller's internel timer to periodically mimic power failure.
Our measurements are not exact, and may over-estimate overheads whose resolutions are finer than what can be precisely
captured by our method, such as Alpaca's logging overhead which is only few instructions.
Nonetheless, we expect the result to show the rough scale of each overhead without deviating too much from the truth.

Figure~\ref{fig:overhead} shows overheads of each system.
The data show that \sys has high performance because it imposes few
overheads. Alpaca's logging requires many fewer operations than \chain's channel
manipulations and Ratchet and \dino's checkpointing. 
Also, the task transition overhead of Alpaca-undo is shown to be much less than the task
transition overhead of Alpaca-redo. This is because Alpaca-redo needs to commit privatized
values on transition, and is the main reason for the speedup of Alpaca-undo against Alpaca-redo.

\subsection{Non-Volatile Memory Consumption}

\begin{wrapfigure}{h!}{0.5\textwidth}
        \setlength{\abovecaptionskip}{3pt}
	\centering
	\includegraphics[width=0.48\textwidth]{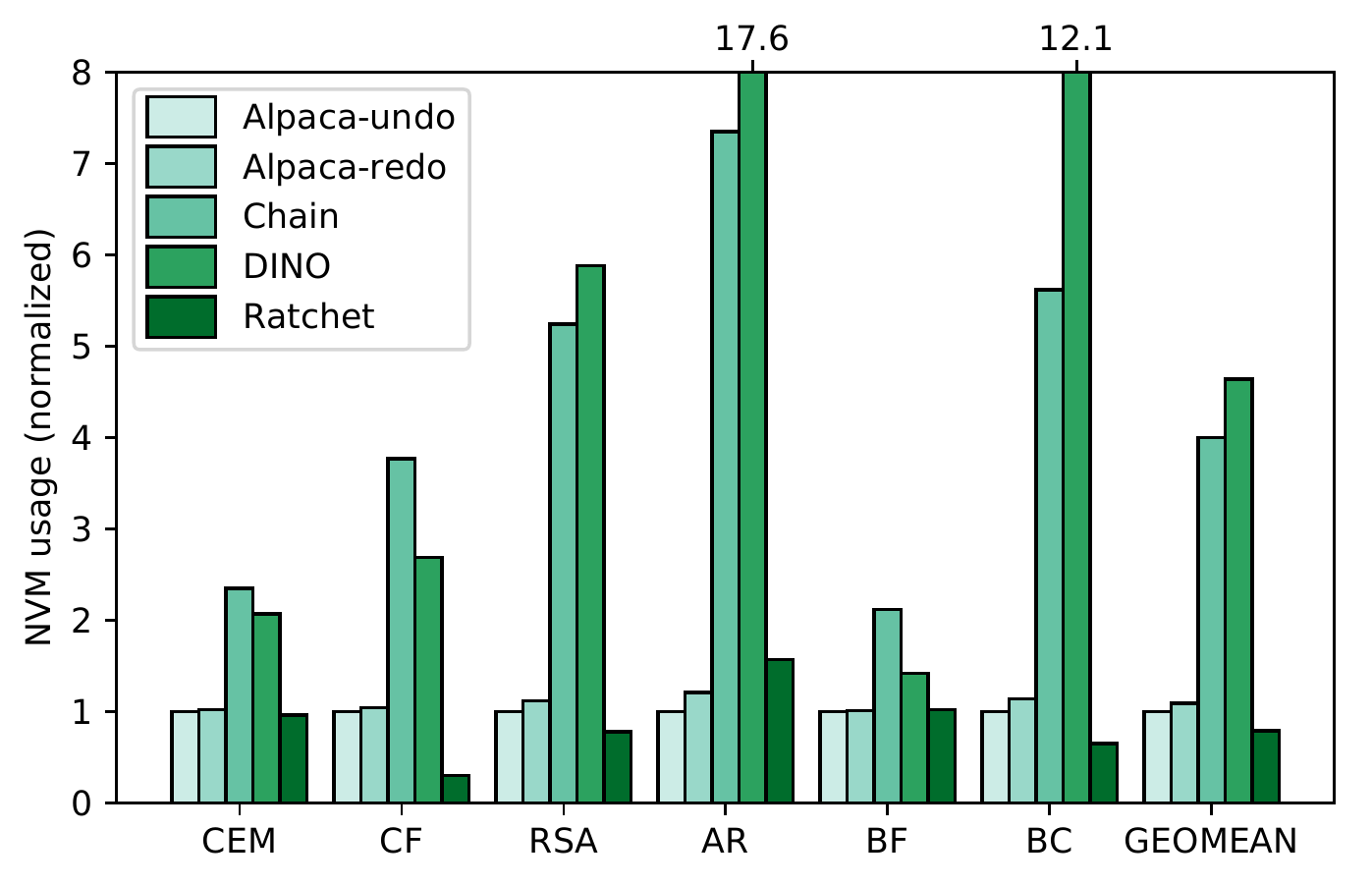}
	\caption{{Non-volatile memory use.}}
	\label{fig:nvm_usage}
\end{wrapfigure}

We measured the non-volatile memory consumption by inspecting each application
binary. For Ratchet which uses FRAM as its main memory, we measured the 
size of the stack to measure the non-volatile memory consumption.
For \dino which reserves double-buffered checkpointing space equal to twice the maximum stack
size of 2KB, i.e., 4KB total, we added 4KB to the number from the binary.
None of these applications dynamically
allocates non-volatile memory, as is typical in embedded systems.   
Figure~\ref{fig:nvm_usage} shows that Alpaca-undo and Alpaca-redo uses
moderate non-volatile memory, using slightly more than Ratchet, but much less
than \chain and \dino.
\sys uses less non-volatile
memory than \chain mainly because \chain creates multiple versions of variables
that exist in different channels. \sys uses less non-volatile memory than \dino
because \dino checkpoints all volatile state and versions some non-volatile
state, while \sys never checkpoints and only selectively privatizes
non-volatile state.

\subsection{Privatizing Data to Volatile Memory}
\label{sec:sysvm}

We evaluated an alternative implementation of Alpaca-redo, called \sys-VM that uses
volatile memory to store privatized values, motivated by the fact that volatile
memory accesses require less energy than non-volatile memory
accesses. Unlike Alpaca-undo whose updates are made in-situ in FRAM, Alpaca-redo privatizes
the \war variable, making privatizing to volatile memory possible.

To ensure that volatile values commit atomically despite failures, \sys-VM must
make a full copy of all privatized {\em values} to a non-volatile commit buffer during
pre-commit.
Privatizing data to volatile memory is only
a net benefit if the time and energy saved by using volatile memory in the task
are more than the time and energy consumed by copying to the commit buffer.

We experimented with a microbenchmark which does fixed number of read-modify-write (RMU)
operation to measure how many accesses to volatile
privatized data are required to amortize the increased pre-commit cost of using
volatile privatization buffers.  
The experiment result implied that when the task contains more than around 110 RMWs,
\sys-VM begins to outperform \sys-redo.

We quantified the number of read and writes per each task in our real applications.
Our tasks had 2.1 reads and 1.05 writes to a privatized variables on average.
The numbers are much smaller than the ``tipping point'' which was around 110 RMWs, 
suggesting that volatile privatization is unlikely to pay off.

We implemented \sys-VM and the result showed that the performance is often worse than,
or negligibly different from \sys-redo's performance, which is consistent with our ``tipping point''
characterization.
\sys-VM is only likely to be viable and
beneficial in a system with a much larger energy buffering capacitor that
accommodates more (i.e., hundreds of) reads and writes in each task. 

\subsection{Comparing Programmer Effort}
\label{sec:prog_effort}

We compared the programming effort required by \sys to the effort required by
\chain, \dino, and Ratchet and found that \sys requires reasonable code changes compared
to Ratchet and DINO code, but requires less change than writing \chain code.  
Like \sys, \chain also requires the programmer to decompose code into tasks,
which is different from writing typical C code and we characterize task sizing
next.  Unlike \sys, \chain also requires additional effort to re-write memory
access code in terms of channel operations, which is different from a typical
C programming.  \sys instead allows code to manipulate task-shared variables
like ordinary C variables using loads and stores.

\subsubsection{Quantifying Programmer Effort}

\begin{table}[h]
\small
    \caption{\label{tbl:effort} {Lines of code and number of keywords.} }
	\begin{tabular}{ l | l | l | l | l | l | l | l | l | l | l}
		{\bf App} & \multicolumn{3}{l|}{{\bf Alpaca}} & \multicolumn{4}{l|}{{\bf Chain}} & \multicolumn{2}{l|}{{\bf DINO}} & {\bf Ratchet} \\ \hline
		{} & {\bf LoC} & {\bf \# Bnd.} & {\bf \# Decl.} & {\bf LoC} & {\bf \# Bnd.} & {\bf \# Decl.} & {\bf \# R/W} & {\bf LoC} & {\bf \# Bnd.} & {\bf LoC} \\ \hline
		{CEM} & {372} & {19} & {28} & {721} & {19} & {40} & {63} & {338} & {13} & {325}\\ 
		{CF} & {397} & {19} & {29} & {707} & {19} & {41} & {72} & {335} & {11} & {324} \\ 
		{RSA} & {765} & {27} & {40} & {1197} & {27} & {53} & {123} & {722} & {35} & {687} \\ 
		{AR} & {466} & {19} & {26} & {713} & {19} & {34} & {57} & {439} & {8} & {431} \\ 
		{BF} & {614} & {18} & {24} & {740} & {18} & {29} & {75} & {556} & {9} & {547} \\ 
		{BC} & {313} & {23} & {26} & {588} & {23} & {26} & {57} & {276} & {10} & {266} \\ 
    \end{tabular}
\end{table}

We quantified the difference in programmer effort between systems by comparing
the differences in the number of lines of code (LoC) and the number of keywords
by each system.  Keywords are divided into three types: boundary, declaration,
and read/write. Boundary keywords (Bnd) represent task boundaries (i.e., \trans) in
\sys and \chain, and checkpoints in \dino. Declaration keywords (Decl) modify function
and data declarations: {\tt task} and {\tt TS} for \sys, and {\tt task} and {\tt
channel} declaration for \chain.  Read/Write keywords (R/W) access memory and 
only occur in \chain (channel in and channel out), because \sys and \dino use
a standard C read/write memory interface.
Ratchet does not require any additional keywords.

Table~\ref{tbl:effort} summarizes the data. On average, the number of lines of
\sys code is 11\% higher than \dino code and 14\% higher than Ratchet, but \sys requires 39\% fewer lines
then \chain. The number of keywords used by \sys code is 240\% more than the
number used by \dino code, but is only 27\% of the number used by \chain code.
Although these data are only a rough indicator of programming complexity, the
data suggest that \sys's complexity lies somewhere between \chain and \dino.

\subsubsection{Choosing a Task's Size}

Dividing a program into tasks is a key part of \sys development, and we we
experimentally characterize the process to show that it is reasonable. 
\sys preserves forward progress at the granularity of a task assuming the
system eventually buffers sufficient energy to complete each task. 
However, a real, energy-harvesting system with a fixed-size energy buffer, may
never be able to buffer sufficient energy for a very long task to complete,
preventing progress. 
If a task is too short, its privatization, commit, and transition overhead will
be relatively very high, impeding performance.  
Based on knowledge of the device and the energy cost of program tasks, the
programmer must assign work to an \sys task.

\begin{figure}
	\centering
	\begin{subfigure}[t]{0.49\textwidth}
		\includegraphics[width=1\textwidth]{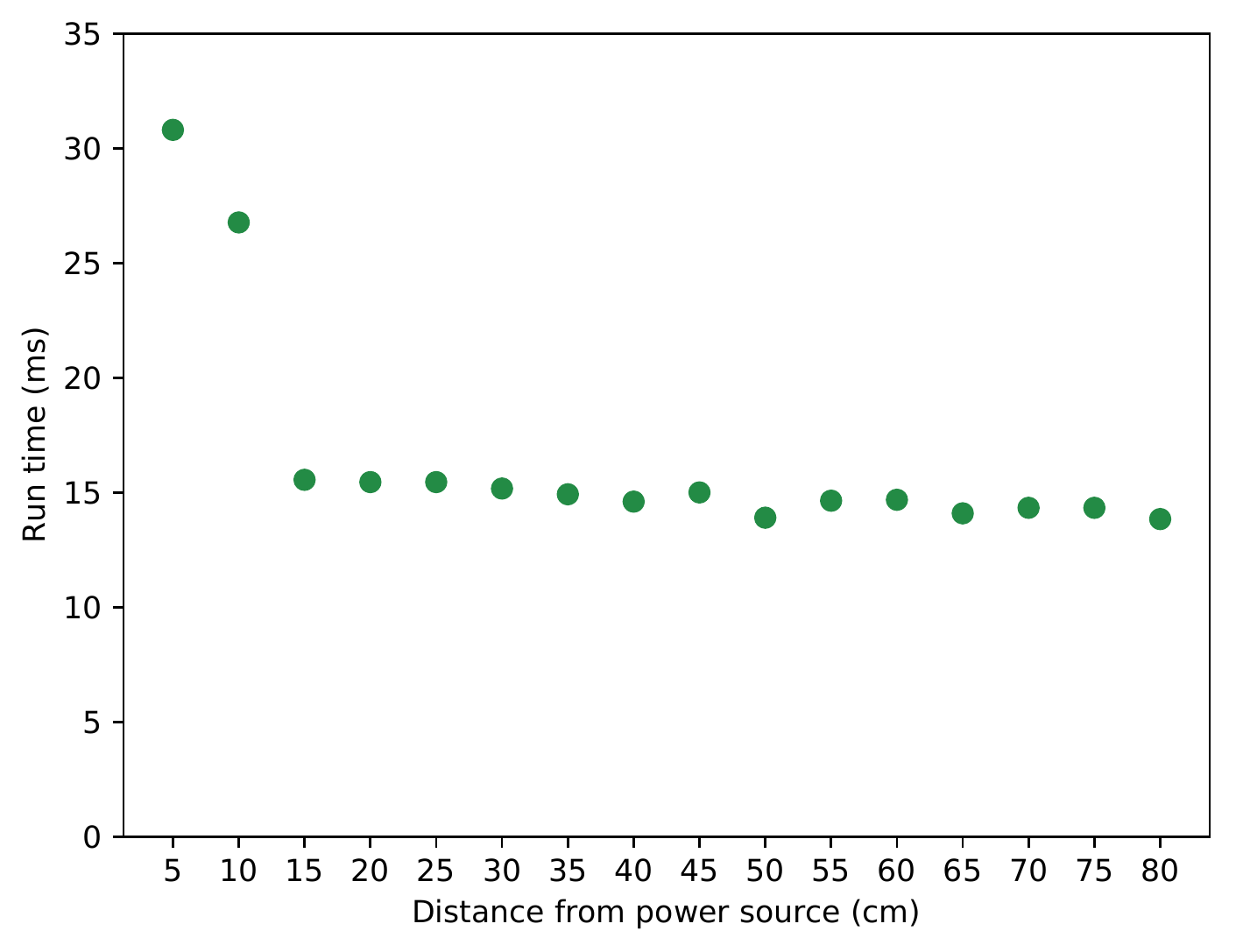}
		\caption{{One charge cycle run time for various distance}}
		\label{fig:runtime_distance}
	\end{subfigure}
	\begin{subfigure}[t]{0.49\textwidth}
		\includegraphics[width=1\textwidth]{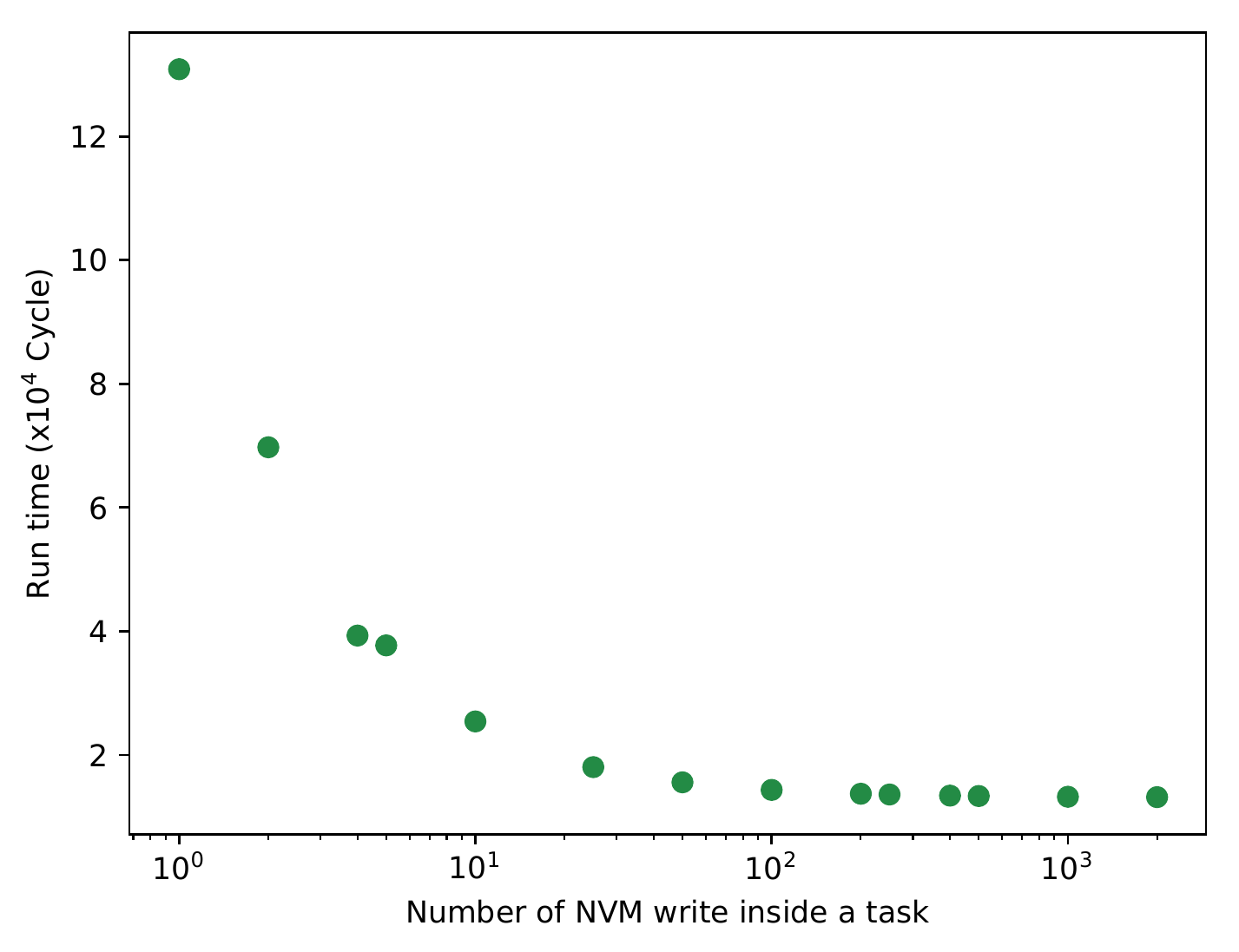}
		\caption{{Run time for various task size}}
		\label{fig:size_runtime}
	\end{subfigure}
	\caption{{Choosing task size.} (a) {\em Energy} availability does not vary with input {\em power}.
	(b) Task overhead varies with task size. }
\end{figure}

While it is a non-trivial programming task, defining the extents of \sys tasks
requires only modest programmer effort.
We observed that on today's energy-harvesting hardware, the task decomposition
problem is independent of input power and depends only on the device's energy
buffer size.
Figure~\ref{fig:runtime_distance} shows data for a microbenchmark that
runs a loop on a WISP5~\cite{wisp} device harvesting energy from an RF
power supply.  The x-axis shows the distance to the RF power
supply, which corresponds to input power.  The y-axis shows the
time to the first brown out, at which point the device has exhausted energy
accumulated in its capacitor and must slowly recharge.  Except for distances so
small that the RF supply effectively continuously powers the device
(\textasciitilde10cm), the amount of work that the system can execute before
browning out is invariant to input power; {\em the energy buffer is constant}. 
Forming \sys tasks is thus a reasonable (albeit non-trivial) task because the
programmer need only reason about the total energy cost of a task. The
programmer need not reason about instantaneous input power, nor the power
envelope of particular hardware operations, which would be
difficult.

We also experimentally observed that choosing a task size that amortizes
privatization, commit, and transition costs is not overly challenging.  On a
WISP5 device, we studied the effect of task size on the run time of a
microbenchmark that executes a fixed amount of work across a varying size of
tasks. The microbenchmark executes a fixed total number of read-modify-write
operations on entries in an array.  We varied the number of accesses per task,
and Figure~\ref{fig:size_runtime} shows the relationship between task size and
total run time.  Run time decreases as task size grows because tasks better
amortize commit and transition cost.  However, the effect saturates as tasks
grow, revealing that even relatively small tasks of around 100
read-modify-write amortize \sys's overheads well.  The data suggest that
choosing a task size that amortizes task overheads will not be prohibitively
challenging to a programmer.

\section{Related Work}
\label{sec:related}

\sys relates to prior work in several areas.  Most related are prior efforts
studying intermittent computing, some of which discussed in
Section~\ref{sec:background}.  We also relate \sys to work on idempotent
compilation, systems with non-volatile memory, transactions and
transactional memory.

\subsection{Energy-Harvesting and Intermittent Computing}
There is a large body of work on intermittent execution and other support for
intermittent systems.  
Some work~\cite{mementos,ratchet,idetic,chinchilla} preserves progress and keep memory
consistent by placing checkpoint automatically that copies the volatile state.
\sys avoids the overhead of volatile state checkpointing and conservatism of the
checkpoint placed by the compiler.

Other work~\cite{dino, chain} versions non-volatile memory either manually~\cite{chain}
or automatically~\cite{dino} to support systems with mixed-volatility~\cite{wisp,moo,mspcdino}.
\sys's non-volatile memory protection is more efficient than the
prior systems (see Section~\ref{sec:eval}).

Some systems~\cite{cleancut, intel, harvos} tries to statically estimate energy
use of a code and optimize checkpoint placement. However, estimating energy use in arbitrary
code is difficult and error prone.
\sys asks the programmer to place the boundaries of the task.

QuickRecall~\cite{quickrecall},
Hibernus~\cite{hibernus}, and Hibernus++~\cite{hibernusplusplus} do on-demand
checkpointing of volatile state when supply voltage is below a threshold.  This
approach is effective, but requires continuous supply voltage measurement
hardware, which is not typically available~\cite{wisp,moo}.  Also, choosing a
threshold voltage is not straightforward. Too high a threshold makes the system
checkpoint and wait for energy, even if there is ample energy to continue. Too
low a threshold may fail to guarantee that checkpointing completes, which is
especially problematic with a variable size call stack and arbitrary global
variables.  \sys is energy agnostic, avoiding hardware requirements and
threshold voltage assignment issues.

Non-volatile processors~\cite{nvp} and Clank~\cite{clank} propose architectural
support making intermittent software simple, but precluding the use of an existing
hardware and imposing a performance and complexity overhead. 
Dewdrop~\cite{dewdrop} runs small, ``one-shot'' tasks on intermittent hardware,
optimizing task scheduling to maximize task completion likelihood given limited
energy.  Dewrop, however, does not support computations that span failures. 

Other work addresses intermittent computation, like \sys, but unlike \sys,
these efforts are not programming or execution models.
Incidental computing~\cite{incidental} and NEOFog~\cite{neofog} optimize specific
applications on top of the non-volatile processor.
Wisent~\cite{wisent,wisentinfocomm17} addresses intermittence, but is not a
computing model, instead enabling reliable software updating of {\em in situ}
intermittent devices. Ekho~\cite{ekho} helps test intermittent devices with
support to collect and replay representative power traces from a realistic
environment.  EDB~\cite{edb} is a hardware/software tool that allows
programmers to profile and debug intermittent devices without interfering with
their energy level.  Federated energy~\cite{ufop} is a disaggregated energy
buffering mechanism that decouples the energy storage of different hardware
components. Flicker~\cite{flicker} eases the design of an energy-harvesting hardware platform by modularized peripherals and harvesters.
TARDIS and CusTARD~\cite{custard} keeps time on power failure and Mayfly~\cite{mayfly}
ensure timeliness of the data.
Capybara~\cite{capybara} enables changing the energy buffer size on-the-fly
to support variety of application demands.
Some earlier work addresses computing using harvested energy, but unlike \sys,
these systems to not explicitly address intermittent computation.
Eon~\cite{eon} is one of the earliest efforts to target harvested-energy
computation, scheduling prioritized tasks based on energy availability.
ZebraNet~\cite{zebranet} dealt with the challenges of solar energy in an
adversarial environment.


\subsection{Idempotent Code Compilation}

Several prior efforts ~\cite{idempotent, idempotent2, conair} noted that a
program decomposed into idempotent sections is robust to a number of failure
modes because idempotent sections can be safely re-executed.  Idempotence
systems break \war dependances by dividing dependent operations with a
checkpoint (or section boundary).  Like these systems, \sys leverages the fact
that eliminating \war dependences makes tasks idempotently re-executable.
Unlike other systems, however, \sys does not make code sections idempotent by
inserting checkpoints. Instead \sys ensures task atomicity by using task-based
execution to avoid the need for volatile state checkpoints, and privatization
of non-volatile data involved in \war dependences to make tasks idempotently
restartable.

As discussed in Section~\ref{sec:background}, \ratchet~\cite{ratchet} uses
compiler idempotence analysis to insert checkpoints to make inter-checkpoint
regions idempotent, assuming main memory is entirely non-volatile. \sys makes
no assumption about memory volatility making it applicable to more varied
hardware, and its tasks' sizes are free from idempotence
analysis, unlike \ratchet.

\subsection{Memory Persistency and Non-Volatile Memory Systems}

The increasing availability of non-volatile memory creates a need for models
defining the allowable reorderings of non-volatile memory updates and {\em
persist} actions, which ensure data become persistent~\cite{pm, pm2}. Relaxing
the ordering of updates and persist actions to different locations may expose a
re-ordering to code resuming execution after a failure and persistency models
describe which of these re-orderings are valid. Other, earlier work developed mechanisms for
managing data structures in non-volatile memory, and for building consistent
memory and file systems out of byte-addressable non-volatile memory~\cite{bpfs,
pmfs, moraru2013, wrap, nvheaps, wholesystempersistence, mnemosyne,cdds}.
\sys relates to these efforts because both aim to keep non-volatile memory
consistent across power failures.  The prior work differs from \sys, however,
in purpose and mechanism. \sys is programming model and run-time implementation
that keeps data consistent across extremely frequent failures in intermittent
executions. These prior efforts focused on large-scale systems and are
only peripherally applicable to intermittent devices.

\subsection{Transactions and Transactional Memory}

Transactions~\cite{gray92} and, in particular, transactional memory~\cite{tm,
stm, tcc, ctm} (TM) systems are also related to \sys. Transactional memory targets
multi-threading systems.  A transaction speculatively updates memory until a
(usually) statically defined atomic region ends. Transactions commit when they
complete execution, updating globally visible state, or aborting their
speculative updates due to a conflicting access in another thread, and
beginning execution again. 
Transactions are similar to \sys because \sys buffers a task's updates
privately, committing them to global memory when a task ends.  Moreover, when a
power failure interrupts a task, its privatized updates are aborted and it
begins again from its start.  However, \sys differs in that it targets
intermittent systems with potentially extremely frequent failures. Unlike TM,
\sys does not target multi-threaded programs, instead aiming to keep
memory consistent between re-executions across power failures.

\section{Conclusion and Future Work}
\label{sec:conclusion} 

This work proposed \sys, a programming model for low-overhead intermittent
computing that does not require checkpointing, using a task-based execution
model and a logging scheme built on idempotence analysis. Compared
to competitive systems from prior work, our \sys prototype
achieves significant performance improvement compared to a variety of systems from the literature.
Looking to the future, \sys emphasizes a need raised by \chain and
\dino for a system to aid, or automate the decomposition of a program into
tasks, which is currently a reasonable task, but mostly a manual process.



\section*{Acknowledgments}

Thanks to the anonymous reviewers for their valuable feedback and to Vignesh Balaji and Emily Ruppel for contributing to discussions about the work.  This work was supported by National Science Foundation Award {\em CNS-1526342}, a Google Faculty Research Award, and a gift from Disney Research. Kiwan Maeng was partially supported by a scholarship from the Korea Foundation for Advanced Studies. Visit {\tt http://intermittent.systems}.                
%

\bibliography{bib}

\end{document}